\documentclass[a4paper,11pt]{article}
\usepackage{rotating}
\usepackage{jheppub} 
\usepackage[utf8]{inputenc}
\usepackage[T1]{fontenc} 
\usepackage{placeins} 
\usepackage{chngpage}
\usepackage{amsmath,amsfonts}
\usepackage{graphicx}
\usepackage{color}
\usepackage{relsize}
\usepackage{epstopdf}
\usepackage{hyperref}
\usepackage{mathrsfs}
\usepackage{ragged2e}
\usepackage{amssymb}
\usepackage{placeins}
\usepackage[normalem]{ ulem }
\usepackage{amsthm}
\usepackage{comment}
\usepackage{graphics}	
\usepackage{caption}
\usepackage{subcaption}
\usepackage{verbatim}
\usepackage{arydshln}

\usepackage{tabularx,booktabs}
\usepackage{multicol}
\usepackage{array,multirow}
\usepackage{booktabs}
\usepackage{colortbl}
\usepackage{float}
\newcommand{\crowcolor}{\rowcolor[rgb]{0.9,0.9,0.9}}
\newcommand{\ctoprule}{\toprule[0.5mm]}
\newcommand{\cbottomrule}{\bottomrule[0.5mm]}

\makeatletter
\gdef\@fpheader{}
\vspace*{0cm}
\makeatother

\begin{document}

\title{Mapping the SMEFT to discoverable models}% Force line breaks with \\

\author[a]{Ricardo Cepedello,}
\author[b]{Fabian Esser,}
\author[b]{Martin Hirsch}
\author[b,c]{and Veronica Sanz}
\affiliation[a]{Institut f\"ur Theoretische Physik und Astrophysik, Universit\"{a}t W\"{u}rzburg, 97074 Würzburg, Germany}
\affiliation[b]{Instituto de F\'isica Corpuscular (IFIC), Universidad de Valencia-CSIC, E-46980 Valencia, Spain}
\affiliation[c]{Department of Physics and Astronomy, University of Sussex, Brighton BN1 9QH, UK}

\emailAdd{ricardo.cepedello@physik.uni-wuerzburg.de}
\emailAdd{esser@ific.uv.es}
\emailAdd{mahirsch@ific.uv.es}
\emailAdd{veronica.sanz@uv.es}

\date{\today}% It is always \today, today,
             %  but any date may be explicitly specified

\abstract{The matching of specific new physics scenarios onto the SMEFT
framework is a well-understood procedure. The inverse problem, the
matching of the SMEFT to UV scenarios, is more difficult and requires
the development of new methods to perform a systematic exploration of
models. In this paper we use a diagrammatic technique to construct in
an automated way a complete set of possible UV models (given certain,
well specified assumptions) that can produce specific groups of SMEFT
operators, and illustrate its use by generating models with no
tree-level contributions to four-fermion (4F) operators. Those
scenarios, which only contribute to 4F at one-loop order, can contain
relatively light particles that could be discovered at the LHC in
direct searches. For this class of models, we find an interesting
interplay between indirect SMEFT and direct searches. We discuss some examples on how this interplay would look
like when combining low-energy observables with the SMEFT
Higgs-fermion analyses and searches for resonance at the LHC. }


\maketitle

\section{Introduction}

In high-energy physics we search for new heavy states through direct
production of resonances at colliders or indirectly via precision
measurements, sensitive to virtual effects from these states. Indirect
searches for new physics are often analysed in the context of an
Effective Field Theory (EFT) approach, and nowadays the most widely
used EFT approach is labelled as SMEFT, or SM EFT \cite{Weinberg:1979sa, Buchmuller:1985jz, Grzadkowski:2010es,
  Lehman:2014jma, Lehman:2015coa, Henning:2015alf, Gripaios:2018zrz,
  Criado:2019ugp, Murphy:2020rsh, Li:2020gnx, Li:2020xlh,
  Liao:2020jmn}.

On the precision frontier, a legacy of LEP $e^+e^-$ collision
measurements, plus other low energy probes such as atomic parity,
provide very stringent constraints on new states, see for example analyses
using low-energy precision data~\cite{Falkowski:2015krw,
  Falkowski:2017pss, Falkowski:2020pma}. Even assuming a minimal
flavour violation structure of couplings, indirect constraints seem to
push the boundary of new physics into the multi-TeV region.

With such high scales, the LHC may have no chance to directly produce
the resonances and could only be sensitive again to indirect
probes. Indeed, a lot of the current focus using SM LHC precision
measurements is placed in SMEFT analyses combining various LHC
channels, as well as legacy measurements at lower energies. Here again
the constraints from global fits including legacy and LHC Run2 data place bounds on the
scale of dimension-six operators in the multi-TeV range, see \cite{Ellis:2020unq} for the most up-to-date analysis using LHC
Higgs, diboson and top data.

The argument that legacy precision measurements point to a high new
physics scale, inaccessible via direct searches, is clearly
scenario-dependent.  Having a closer look to this issue, one notices
that the tightest SMEFT bounds come from a particular set of EFT
operators, namely those involving four fermions (4F), particularly
those 4F with at least some leptons. Indeed, constraints on
flavour-independent 4F operators range from percent to permille
level~\cite{Falkowski:2017pss}. (We note that it is well-known that
constraints on generation off-diagonal entries in the 4F operators are
even more stringent, in particular for operators containing 
leptons \cite{Crivellin:2017rmk}. Thus, in all of this paper we will 
consider only the generation diagonal parts of the operator, as is 
usually done in LHC analysis.) Therefore, new resonances producing 4F
operators at tree-level are constrained to mass regions of the order
of $m > \lambda^2 \, \times$ (multi-TeV), with $\lambda$ the coupling of the
state to the SM fermions. 

This conclusion is correct, of course, only if the new states produce
the 4F operators at tree-level. On the other hand, in scenarios where
4F are loop-induced at leading order the constraints would be reduced
by a factor of order $1/16\pi^2$. Then the new resonances could be
much lighter, directly accessible at colliders. In this class of scenarios, information from low-energy precision measurements and
collider searches would be complementary.

A systematic description of such {\it discoverable} scenarios is the
focus of this paper. Now, given a particular non-renormalisable
operator, it is in principle straightforward to find some particular
field (or fields) beyond the Standard Model particle content that will
generate the operator in the EFT limit. Due to the finite number of
operators in SMEFT at $d=6$, it is even possible to find the complete
list of such ultra-violet, renormalisable completions ``by hand'', if
one restricts the search to tree-level diagrams
\cite{deBlas:2017xtg}.\footnote{Some attempts on constructing UV complete models for the SMEFT at $d=7$ and $d=8$ have been made in \cite{Li:2022abx}, although one should note this work is not fully consistent with previous works, e.g. Ref.~\cite{Bonnet:2009ej}.}However, already at one-loop level the number of
possible ultra-violet completions (``models'') increases so rapidly
that for any attempt to systematically classify them,
automation becomes mandatory.

The strategy we use in this paper for this task is based on a
diagrammatic method. We will discuss the basics of this method in
section~\ref{sect:expl}.  The same method was used previously in the
deconstruction of the $d=5$ Weinberg operator,
see~\cite{Bonnet:2012kz,Sierra:2014rxa} and in particular
\cite{Cepedello:2018rfh}. Note that our method for operator
deconstruction overlaps in considerable parts with the one discussed
recently in \cite{DasBakshi:2021xbl,Naskar:2022rpg,Banerjee:2020bym}. The diagrammatic method can
be formulated in a general way, such that one can, in principle,
``deconstruct'' or ``explode''~\footnote{Finding the UV completion for
  operators has been called ``exploding'' the operators in
  \cite{Gargalionis:2020xvt}.}  any EFT operator at arbitrary
dimension. For reasons discussed above, however, in this paper we
concentrate on 4F operators in SMEFT only.
\\

The paper is organised as follows: In section~\ref{sect:expl} we present
the main ideas of the diagrammatic method we have developed to
systematically find loop-induced 4F operators and describe conditions
to further restrict the UV scenarios. In
section~\ref{sec:model_candidates} we count the number of UV completions,
describe common patterns among these models, propose a benchmark for
analysing the phenomenology and explore scenarios that lead to 4F with
just quarks, see also Appendix~\ref{App:quark_models}.  In
section~\ref{sec:pheno} we use the model patterns to describe the
typical phenomenology at the LHC and give specific predictions for our
benchmark model. We finish by presenting our conclusions in
section~\ref{sec:concls}.

%%%%%%%%%%%%%%%%%%%%%%%%%%%%%%%%%%
%%%%%%%%%%%%%%%%%%%%%%%%%%%%%%%%%%
\section{One-loop models for SMEFT 4F operators\label{sect:expl}}

In this section we discuss the construction of one-loop models for SMEFT 4F operators. The well-known list of baryon and lepton
number conserving operators in the Warsaw basis is given in
table~\ref{tab:4-fermion_operators}.  As mentioned above, while at
tree-level there is a small and fixed number of UV completions, at
one-loop level a huge number of valid models can be constructed. In
subsection \ref{subsec:diag} we briefly describe how we construct
the list of models in an automated way. Since in this work we are
mainly interested in models testable at the LHC, we will apply a
number of restrictions on the list of models. These will be discussed
in subsection \ref{sec:UV_model_constraints}.

%%%%%%%%%%%%%%%%
\begin{table}[ht!]
    \centering
    \begin{tabular}{|l|l|l|}
        \hline
        Class & Name & Structure \\
        \hline \hline
        LL & $\mathcal{O}_{ll}$ &  $ \left(\bar{l}_L \gamma_{\mu} l_L\right) \left(\bar{l}_L \gamma^{\mu} l_L\right)$ \\
        \hline
        & $\mathcal{O}_{le}$ & $ \left(\bar{l}_L \gamma_{\mu} l_L \right)\left(\bar{e}_R \gamma^{\mu} e_R \right)$ \\
        \hline
        & $\mathcal{O}_{ee}$ & $ \left(\bar{e}_R \gamma_{\mu} e_R \right) \left(\bar{e}_R \gamma^{\mu} e_R \right)$ \\
        \hline \hline
        LQ & $\mathcal{O}_{lq}^{(1)}$ & $\left(\bar{l}_L \gamma_{\mu} l_L\right) \left(\bar{q}_L \gamma^{\mu} q_L\right)$ \\
        &$\mathcal{O}_{lq}^{(3)}$ &  $\left(\bar{l}_L \gamma_{\mu} \sigma_a l_L\right) \left(\bar{q}_L \gamma^{\mu} \sigma_a q_L\right)$ \\
        \hline
        & $O_{lu}$ & $ \left(\bar{l}_L \gamma_{\mu} l_L\right) \left(\bar{u}_R \gamma^{\mu} u_R\right)$ \\
        \hline
        &  $\mathcal{O}_{ld}$ & $\left(\bar{l}_L \gamma_{\mu} l_L\right)\left(\bar{d}_R \gamma^{\mu} d_R\right) $ \\
        \hline
        & $\mathcal{O}_{lequ}^{(1)}$ &  $\left(\bar{l}_L e_R \right) i \sigma_2 \left(\bar{q}_L u_R \right)^T$ \\
        & $\mathcal{O}_{lequ}^{(3)}$ & $\left(\bar{l}_L \sigma_{\mu\nu} e_R \right) i \sigma_2 \left(\bar{q}_L \sigma^{\mu \nu} u_R \right)^T$ \\
        \hline
        & $\mathcal{O}_{ledq}$ & $\left(\bar{l}_L e_R \right) \left(\bar{d}_R q_L \right)$ \\
        \hline
        & $\mathcal{O}_{qe}$ & $\left(\bar{q}_L \gamma_{\mu} q_L\right) \left(\bar{e}_R \gamma^{\mu} e_R\right)$ \\
        \hline
        & $\mathcal{O}_{eu}$ & $\left(\bar{e}_R \gamma_{\mu} e_R\right) \left(\bar{u}_R \gamma^{\mu} u_R\right)$ \\
        \hline
        & $\mathcal{O}_{ed}$ & $\left(\bar{e}_R \gamma_{\mu} e_R\right) \left(\bar{d}_R \gamma^{\mu} d_R\right)$ \\
        \hline \hline
        QQ & $\mathcal{O}_{qq}^{(1)}$ & $\left(\bar{q}_L \gamma_{\mu} q_L\right) \left(\bar{q}_L \gamma^{\mu} q_L\right)$ \\
        &  $\mathcal{O}_{qq}^{(3)}$ & $\left(\bar{q}_L \gamma_{\mu} \sigma_a q_L\right) \left(\bar{q}_L \gamma^{\mu} \sigma_a q_L\right)$ \\
        \hline 
        & $\mathcal{O}_{quqd}^{(1)}$ & $\left(\bar{q}_L u_R \right) \left(\bar{q}_L d_R \right)^T$ \\
        & $\mathcal{O}_{quqd}^{(8)}$ & $\left(\bar{q}_L T_A u_R \right) \left(\bar{q}_L T_A d_R \right)^T$ \\
        \hline
        & $\mathcal{O}_{qu}^{(1)}$ & $\left(\bar{q}_L \gamma_{\mu} q_L\right) \left(\bar{u}_R \gamma^{\mu} u_R\right)$ \\
        & $\mathcal{O}_{qu}^{(8)}$ & $\left(\bar{q}_L \gamma_{\mu} T_A q_L\right) \left(\bar{u}_R \gamma^{\mu} T_A u_R\right)$ \\
        \hline 
        & $\mathcal{O}_{qd}^{(1)}$ & $\left(\bar{q}_L \gamma_{\mu} q_L\right)\left(\bar{d}_R \gamma^{\mu} d_R\right)$ \\
        & $\mathcal{O}_{qd}^{(8)}$ & $\left(\bar{q}_L \gamma_{\mu} T_A q_L\right) \left(\bar{d}_R \gamma^{\mu} T_A d_R\right)$ \\
        \hline
        & $\mathcal{O}_{uu}$ & $ \left(\bar{u}_R \gamma_{\mu} u_R\right) \left(\bar{u}_R \gamma^{\mu} u_R\right)$ \\
        \hline
        & $\mathcal{O}_{ud}^{(1)}$ & $\left(\bar{u}_R \gamma_{\mu} u_R\right) \left(\bar{d}_R \gamma^{\mu} d_R\right)$ \\
        & $\mathcal{O}_{ud}^{(8)}$ & $\left(\bar{u}_R \gamma_{\mu} T_A u_R\right) \left(\bar{d}_R \gamma^{\mu} T_A d_R\right)$ \\
        \hline
        & $\mathcal{O}_{dd}$ & $\left(\bar{d}_R \gamma_{\mu} d_R\right) \left(\bar{d}_R \gamma^{\mu} d_R\right)$ \\
        \hline
    \end{tabular}
    \caption{List of baryon (and lepton) number conserving
      4F operators in the Warsaw basis at $d =$6. Note that
      we have suppressed generation indices here.}
    \label{tab:4-fermion_operators}
\end{table}
%%%%%%%%%%%%%%%%%%%%% TABLE

\subsection{From operators to models\label{subsec:diag}}

Here, we give a brief summary of the diagrammatic method, 
for further details see \cite{Cepedello:2018rfh}. The method involves 
essentially three steps. For any given operator, one can first find 
all topologies with $n$ external legs and $k$ loops. Since the aim is to 
find renormalisable UV models, only topologies with 3- and 4-point vertices
are kept. Into these topologies one then inserts fermions, scalars
and vectors in all possible ways allowed by Lorentz invariance to
find the complete list of diagrams. 

\begin{figure}[t]
\begin{center}
\includegraphics[width=1.\linewidth]{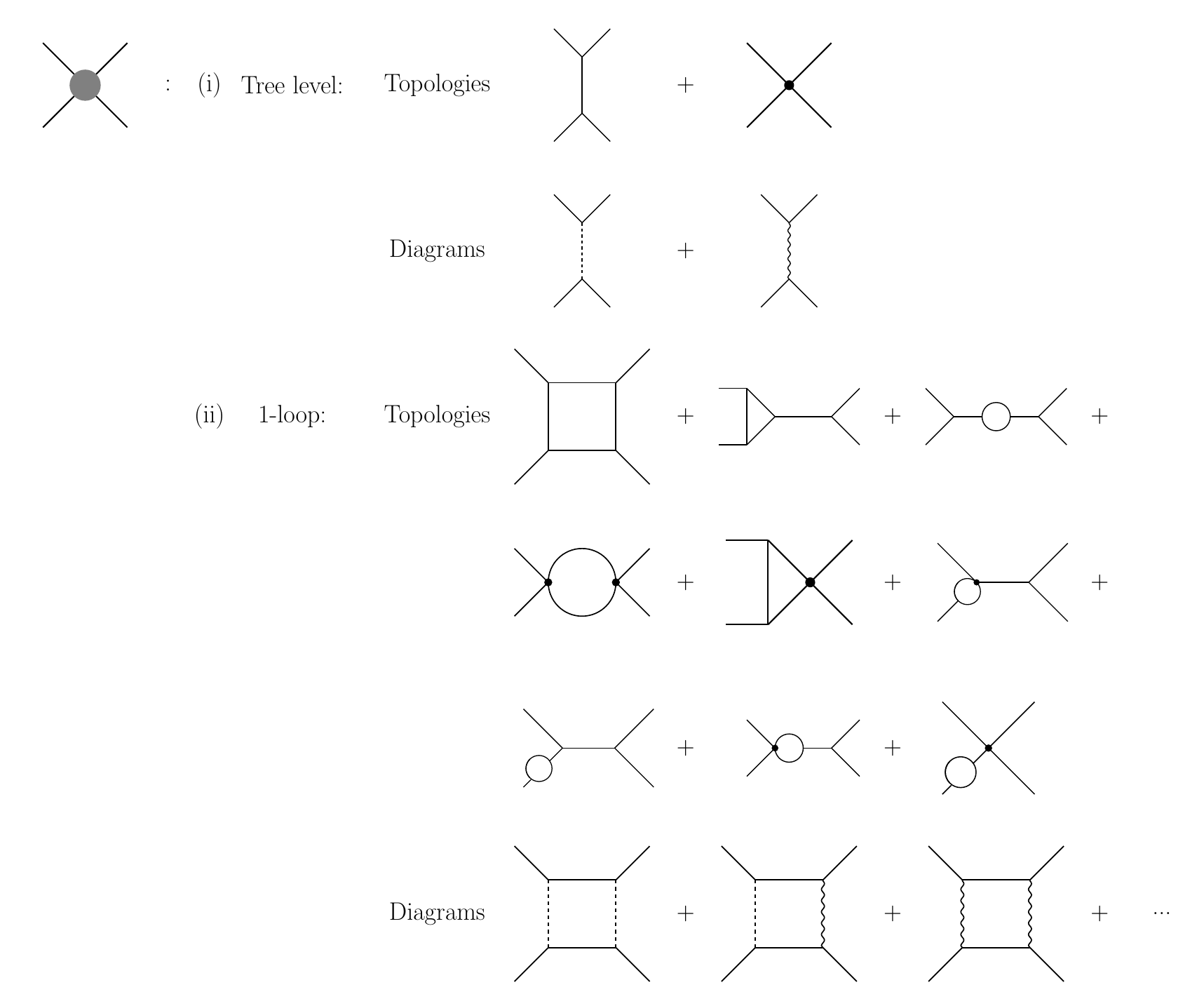}
\caption{The figure illustrates the first two steps in the diagrammatic method for the construction of UV model lists. 4-point vertices are marked with a dot. For a discussion 
see text.}
\label{fig:TopoDiag}
\end{center}
\end{figure}

Figure~\ref{fig:TopoDiag} outlines how the procedure works
schematically for the example of 4F operators. The upper
half of the figure shows the tree-level case. For this simple operator
at tree-level there are only two topologies. The second of these
involves a single four point vertex, and thus, for fermions, is a
non-renormalisable vertex. This topology is therefore immediately
discarded as we are interested in UV renormalisable models. The first topology can be realised by only two diagrams,
which are shown below the topologies. At one-loop level, there are already nine
topologies, but again five of those can be thrown away immediately in the
case of a 4F operator since they contain 4-point vertices
connected to the outside of the topology. Of the remaining four
topologies, the one in the last line of topologies is a
self-energy that only renormalises the propagator of one of the
outside fermions. We will not discuss this topology further here, see
the discussion on matching below.  This leaves the three topologies in
the first line for further study. As in the case of the tree-level,
one can now insert fermions, scalars and vectors into the topologies
in all physically consistent ways. Note that the figure shows only a
few diagrams for the box topology, the list of diagrams shown in this
figure is not complete at one-loop order. Both steps can be automatised
easily, since mathematically topologies and diagrams can be
represented in terms of adjacency matrices. In these adjacency matrices rows and columns represent each vertex, so they should be renormalisable and allowed by Lorentz invariance. We have implemented this procedure in \verb|Mathematica|.

\begin{figure}[t]
\begin{center}
\includegraphics[width=1.\linewidth]{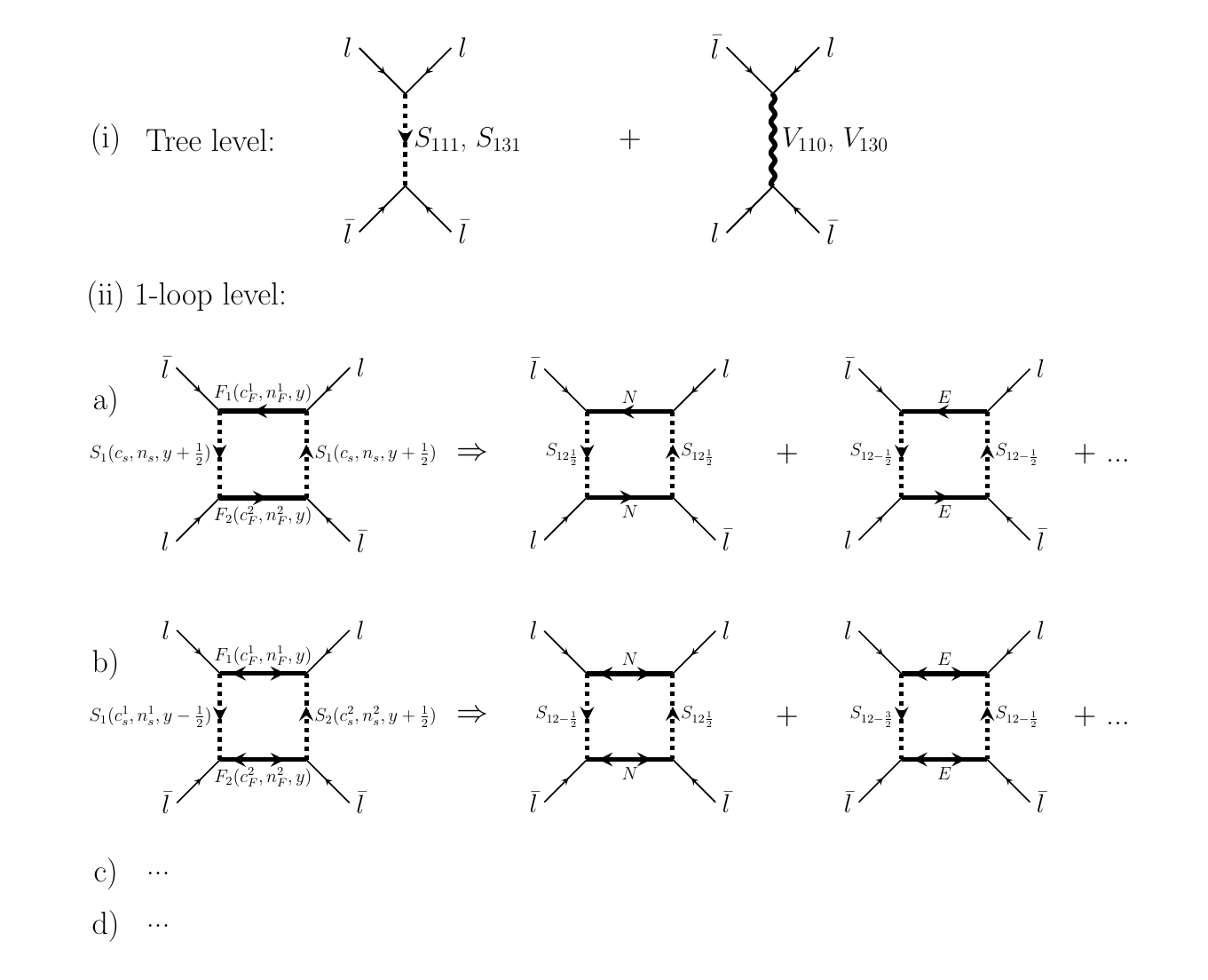}
\caption{The figure illustrates the construction of UV models for a
  given diagram with fixed outside legs. For simplicity of
  presentation we have chosen a symmetric operator, ${\cal O}_{ll}$
  in this example, since it yields the minimal number of models (four
  at tree-level). Note that in diagrams with specific particles
  inserted, we use arrows to track the flow of quantum numbers, such as colour and hypercharge. For vector-like and Majorana fermions, the double arrows indicate a mass
  insertion.}
\label{fig:FigModel}
\end{center}
\end{figure}

In the final step one then needs to identify the quantum numbers for
the internal fields of the diagram, given a set of external legs. The
procedure is depicted schematically in 
figure~\ref{fig:FigModel} for
the example of ${\cal O}_{ll}$. Consider for simplicity first the
tree-level:  Because ${\cal O}_{ll}$ is symmetric in the fields, apart
from generation indices, there are only two possible combinations for
the outside fields. Both lead to exactly two ``models'' (due to the
$SU(2)$ rule ${\bf 2} \otimes {\bf 2} = {\bf 3} \oplus {\bf 1}$).
Note that the ${\bf 1}$ is anti-symmetric and thus vanishes for a
single generation of leptons.

At one-loop level, there is the complication that at the four corners of the
loop diagram one SM field couples to two unknown fields, thus the quantum
numbers of the particles in the loop are not unequivocally fixed. We
can write the quantum numbers of the scalars and fermions in the
diagrams as $S_i(c_S^i,n_S^i,y_S^i)$ and $F_i(c_F^i,n_F^i,y_F^i)$,
with $i=1,2$ (in some examples only one type of fermion and/or scalar
appears) and so on. For this simple 4F operator $c_S^1=c_F^1$ and
$n_F^1=n_S^1 \pm 1$, while $y^i_S$ and $y_F^i$ are related by $\pm
1/2$. For other operators the relations may be more complicated, but
with the help of, for example, \verb|GroupMath| \cite{Fonseca:2020vke}
any necessary set of rules can be calculated in an automated way. A
list of valid models can then be arrived at by inserting a set of
seed fields. The list of seeds needs to fix one of the particles
internal to the loop, all other particle quantum numbers can then be
calculated. The conditions that we use in the current paper to select the
lists of seed fields (and all other cuts on the models) are discussed in subsection
\ref{sec:UV_model_constraints}.  The method itself, however, is
completely agnostic to this choice and works for any list of seeds.
Finally, this procedure has to be repeated for all valid permutations
of outside fermion fields. The result is a list of adjacency matrices,
whose entries are the quantum numbers and Lorentz nature of the
particles in the diagram. From this list one finally deletes all model
duplicates.\footnote{As a technical aside, we add that we keep track
  of the fermion chiralities in the diagrams too, since this determines
  the power of loop momenta as well as possible internal SM fields. 
  For example, boxes with an odd power of
  loop momenta vanish identically for operators without derivatives
  (such as the 4F under consideration here) and thus such diagrams do
  not represent valid models for 4F operators.}

We call diagrams with all particle properties inserted (both Lorentz 
nature and quantum numbers) ``model-diagrams'' or models for short. 
Strictly speaking, the diagrammatic method, as described above, does
not give full-fledged models, it only specifies their particle
content. However, again existing programs such as \verb|Sym2Int|
\cite{Fonseca:2017lem} make it possible to calculate the full
Lagrangian of these models with only the particle content as input in
an automated way.

\begin{figure}[t]
\begin{center}
\includegraphics[width=1.\linewidth]{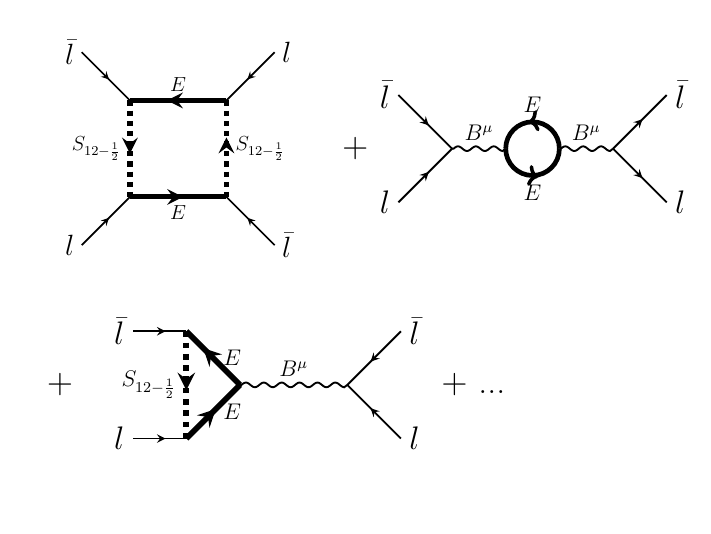}
\caption{The figure illustrates how light particle reducible 
diagrams are generated, going from the complete Green's 
basis to the on-shell Warsaw basis, see text.\label{fig:FigMatch}}
\end{center}
\end{figure}

While the diagrammatic method can be used for model constructions for
any type of operator, there is a subtlety involved in the matching of
models to a particular operator basis.  Let us discuss this in some
detail. The Warsaw basis for $d=6$ SMEFT \cite{Grzadkowski:2010es}
gives the complete on-shell list of operators. It uses all possible
relations among operators based on integration-by-parts (IBP),
equations of motion (EOM) and Fierz transformations to eliminate
redundant operators. While it obviously makes sense physically not to
list redundant operators, it is sometimes very useful to consider the
over-complete set of operators in the so-called Green's basis instead.

Consider, for example, $(D_{\mu}X^{\mu\nu})^2$, where $X^{\mu\nu}$
stands symbolically for any of the three SM fields strength tensors.
This operator exists in the Green's basis, but is eliminated in the
Warsaw basis by the equation of motion for the field strength tensor:
${\cal O}_{D^2X^2} \propto (D_{\mu}X^{\mu\nu}) ={\bar\psi}\gamma^{\nu}\psi 
+\phi^\dagger i {\overleftrightarrow D^{\nu}}\phi$, where $\psi$ includes 
any SM fermion charged under the group to which $X^{\mu\nu}$ belongs.
Any BSM state that is not a complete singlet under the SM group will
couple also to gauge bosons.  Thus, any non-singlet BSM fermion will
appear in ${\cal O}_{D^2X^2}$ at one-loop level with a contribution
proportional to $g_i^2$, which then gives a contribution to all 4F operators involving $\psi$ coupled to $X^{\mu\nu}$, when going
to the on-shell Warsaw basis.

In the diagrammatic method, when fixed to the Warsaw basis, these
contributions appear as light-particle reducible diagrams contributing
to the operator under consideration, see figure~\ref{fig:FigMatch}. The
figure shows for one example model how not only the box diagrams
contribute to the operator ${\cal O}_{ll}$ and lists the other
diagrams in which $B^{\mu\nu}$ appears. (Again, the list of diagrams
shown is not complete. The figure is for illustration only.) Compare
to the topologies in the lower half of figure~\ref{fig:TopoDiag}. Thus,
changing between Green's and Warsaw basis affects the matching of
4F operators in a non-trivial way and one always needs to
specify the basis, in which the matching results are quoted.
The construction of model diagrams is simpler in the Green's basis, since
there we do not need to consider light-particle reducible diagrams,
while in the Warsaw basis (with fewer operators) such diagrams need to
be included for a complete matching since they correspond to applying
EOMs to the operators in the Green's basis.

Note, however, that contributions due to $(D_{\mu}X^{\mu\nu})^2$ are
universal and diagonal in flavour space.  Matching 4F
operators at one-loop will contain terms proportional to, say,
$\lambda^4$, $g_i^2\lambda^2$, $\lambda^2\lambda_{SM}^2$ etc., where
$\lambda$ stands symbolically for a BSM Yukawa, $\lambda_{SM}$ are SM
Yukawas and $g_i$ is any of the gauge couplings.  Since $\forall
\lambda_{SM}\ll 1$ (except for the top quark) and also $g_1,g_2<1$
while there is a priori no reason why $\lambda$ should be small, for a
rough estimate it is often sufficient to discuss matching in the
limit where all SM couplings are set to zero. Diagrammatically this
corresponds to considering only box diagram contributions to the
4F operators.  For the example models discussed in more
detail in section \ref{sec:EU}, however, we also briefly comment on
the complete matching.

The output of our method is a list of model diagrams. For the matching
of specific models onto the SMEFT operators we use
\verb|MatchmakerEFT| \cite{Carmona:2021xtq}.  Note that
\verb|MatchmakerEFT|  follows a similar logic in the calculation of
matching coefficients as discussed here. It
calculates matching first in the Green's basis before converting the
output to the Warsaw basis (or any other basis provided by the user).

\subsection{Restrictions for LHC-testable UV 
models\label{sec:UV_model_constraints}}

In this subsection, we discuss the three main restrictions which we
apply to our list of models. As a short summary up front: (a) Use only models
with BSM scalars and fermions, but no new vectors; (b) select only
models which do not produce any stable charged relics; and (c) select
only models for which no 4F operator appears at tree-level.

Let us discuss vectors first.  Ref.~\cite{deBlas:2017xtg} gives the
complete lists of BSM particles, including also all vectors,  which
could contribute to $d=6$ SMEFT at tree-level. However, while it is
fairly straightforward to construct all diagrams containing vector
particles, consistent model building for {\em gauge vectors} is highly
non-trivial.

There are two unrelated problems for models with gauge vectors.
First, gauge vectors are in the adjoint of the gauge group.  From the
quantum numbers of the vectors one can find the minimal gauge group
in which a given vector can be part of the adjoint. This exercise has
been done, for example, in \cite{Fonseca:2016jbm}. Note that that
paper studied neutrinoless double beta decay, but the list of vectors
covers all but one of the vectors listed in~\cite{deBlas:2017xtg}.
\footnote{\cite{Fonseca:2016jbm} contains more vectors than the list
  in \cite{deBlas:2017xtg}, since it discusses double beta decay up to
  $d=11$. The one missing vector, on the other hand, is 
${\cal U}_5=V_{3,1,5/3}$. This vector appeared, for example, in
  \cite{Biggio:2016wyy}, where it was found that it could be generated
  from the Pati-Salam group, albeit in a non-standard embedding.}  It
has been found that for the majority of these vectors, the groups are very
exotic and it is not even possible to construct a model which can be
broken to the SM gauge group with SM field content in a consistent
way~\cite{Fonseca:2016jbm} if the SM fermions transform non-trivially
under the extended gauge group too.

The second problem with vectors is purely phenomenological. For a
number of vectors in the list of ``exit'' vectors one finds that the
minimal possible gauge groups are grand unified groups, such as
$SU(5)$, flipped $SU(5)$ or also $SO(2n)$ groups. In these cases,
proton decay constraints require that the vectors live at the GUT
scale, rendering them completely uninteresting for electro-weak scale
phenomenology. We note that the recent work \cite{Fonseca:2022wtz}
circumvents this problem partially by assuming all SM fermions to be
singlet under the new gauge group. Interactions with the new gauge
bosons are then generated via mixing of the SM fermions with exotic,
vector-like copies, where the latter are assumed to transform
non-trivially under the new gauge group {\em and} the SM
group. However, while in such a setup the exotic gauge bosons can be
lighter than the classical GUT scale in principle, the model 
construction has to be done in a specific way, such that baryon and lepton 
numbers are conserved, otherwise the proton decay constraints will 
apply again. Given these complications, in this paper we will
concentrate exclusively on models with only scalar and fermion BSM fields.

The second restriction we use to select models is purely
phenomenological.  Consider the following: At one-loop level vertices
will appear in which the BSM fields couple in pairs to a single SM
particle. The simplest possible example is a gauge boson coupling to a
pair of fermions $\psi-{\bar \psi}$. Since one can always find a
singlet in $\psi-{\bar \psi}$, it seems one could build an {\em
  infinite tower of one-loop models}, simply by going to larger
representations and/or larger hypercharges. However, in practice, there
are restrictions for model building, which will provide a cutoff
in this a priori infinite series. 

For model constructions which aim at generating interesting LHC
phenomenology, the most important restriction is that experimental
searches for stable charged relics essentially exclude the mass range
$M \sim [1, 10^5]$ GeV, see for example
\cite{ParticleDataGroup:2020ssz,Hemmick:1989ns,Kudo:2001ie,Taoso:2007qk}.
Thus no model containing a stable, charged BSM field can give any LHC
phenomenology. To avoid this constraint we have to assume either (a)
the BSM fields are unstable or (b) the lightest of the BSM fields is a
good dark matter candidate.  These two possibilities will lead to very
different phenomenology at the LHC, if the BSM particles are light
enough to be produced on-shell. Here, we will concentrate on models in
class (a). However, we plan to return to the dark matter option in a
future publication.

Consider class (a). In the following, we will call certain BSM fields
``exit particles'', or exits for short.  An exit is defined as any BSM
particle that can appear linearly in a Lagrangian term with SM
fields. Such particles then are guaranteed to be unstable. They
are therefore synonymous with the list of particles that can generate
the $d=6$ SMEFT operators at tree-level. The complete lists of these
exits can be found in~\cite{deBlas:2017xtg}. As already mentioned in
the introduction, this list is finite. Thus, at one-loop level also the
number of models pertaining to this class is finite.

Before turning to our third restriction on model building, we mention
in passing that also other arguments providing a cutoff for the
``infinite'' series of one-loop models exist. For example, one may want
to argue that any interesting BSM extension should maintain the
perturbativity of the gauge couplings up to some large energy scale,
say for example the grand unification scale, $m_G$ or even the Planck
scale. See for example \cite{Arbelaez:2022ejo} for a recent
discussion on how this argument limits the number of viable models in
the case of the Weinberg operator. In this paper, we do not use
such theoretical arguments. 

\begin{figure}
    \begin{subfigure}[b]{0.5\textwidth}
       \centering
       \includegraphics[width=0.7\textwidth]{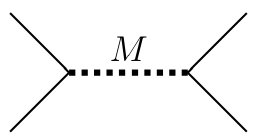}
	\subcaption{At tree-level, $c_{4F} \propto \frac{1}{M^2}$}
    \end{subfigure}
    \begin{subfigure}[b]{0.5\textwidth}
	    \centering
	    \includegraphics[width=0.7\textwidth]{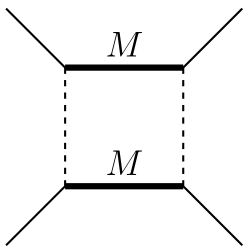}
	\subcaption{At one-loop level, $c_{4F} \propto \frac{1}{16 \pi^2} \frac{1}{M^2}$}
    \end{subfigure}
    \caption{Sample diagrams at tree-level and one-loop level. The one-loop
      contribution to the Wilson coefficient is suppressed by a factor
      of $16\pi^2$ with respect to the tree-level contribution.}
    \label{fig:interplay_tree_1loop}
\end{figure}

Finally, we will further limit our list of models by excluding all
models of a one-loop generated 4F operator whose particle
content produces any other 4F operator at
tree-level. Again, this restriction is simply driven by our desire to
find models that can be studied at the LHC. Tree-level generated $d=6$
operators will lead to restrictions on the new physics scale
$\Lambda$ which are typically stronger by a factor of $4\pi$ than the
ones found for one-loop generated operators as discussed above, see
also Fig.~\ref{fig:interplay_tree_1loop}.

In terms of model generation, the practical consequence of this choice
is that not all possible scalar exits can appear in our model list.
In Table \ref{t:scalars} we list the four scalar exits in this
class. ${\cal S}$ and $\Xi$ will decay to two Higgses (or, of course,
gauge bosons, after electro-weak symmetry breaking), while
$\Theta_{1,3}$ decay to three Higgses.  Table \ref{t:fermions} gives
the list of fermion exits for completeness. None of the fermions can
produce a 4F operator at tree-level, thus this list is
identical to the one found in \cite{deBlas:2017xtg}.

One last, purely phenomenological, restriction we use to limit the
number of generated models consists in neglecting all models that
produce only diagrams that are proportional to at least some SM Yukawa
($\lambda_{SM}$) or gauge ($g_i$) couplings, for example $\lambda_{SM}^2\lambda^2$.  The
reason for not considering such models is that, except for the top
quark, $\lambda_{SM} \ll 1$ and consequently a diagram involving an
SM Yukawa coupling will be suppressed with respect to any diagram that
contains only BSM couplings that we assume to be naturally of order
${\cal O}(1)$. However, as noted above, in the actual matching process
done by \verb|MatchmakerEFT| all SM Yukawa couplings will be taken into
account, leading to ``exact'' matching results.

\begin{table}[t]
  \begin{center}
    {\small
      \begin{tabular}{lcccc}
        \ctoprule
        \crowcolor
        Name &
        ${\cal S}$ &
        $\Xi$ &
        $\Theta_1$ &
        $\Theta_3$ \\
        Irrep &
        $\left(1,1,0\right)$ &
        $\left(1,3,0\right)$ &
        $\left(1,4,{\frac 12}\right)$ &
        $\left(1,4,{\frac 32}\right)$ \\[1.3mm]
        \cbottomrule
      \end{tabular}
    }
    \caption{Scalar ``exits'' that do not generate tree-level
      four-fermion operators. Symbols use the notation of \cite{deBlas:2017xtg}. The numbers in brackets refer to the representations of the SM gauge groups $SU(3)_C$, $SU(2)_L$ and $U(1)_Y$.}
    \label{t:scalars}
  \end{center}
\end{table}

%
%%%%%%%%%%%%%%%%%%%%%%%%%%%%%%
% TABLE FERMIONS
%
\begin{table}[t]
  \begin{center}
    {\small
      \begin{tabular}{lccccccc}
        \ctoprule
        \crowcolor
        Name &
     $N$ & $E$ & $\Delta_1$ & $\Delta_3$ & $\Sigma$ & $\Sigma_1$ & \\
        Irrep &
        $\left(1, 1,0\right)$ &
        $\left(1, 1,-1\right)$ &
        $\left(1, 2,-\frac{1}{2}\right)$ &
        $\left(1, 2,-\frac{3}{2}\right)$ &
        $\left(1, 3,0\right)$ &
        $\left(1, 3,-1\right)$ & \\[1.3mm]
       \cbottomrule
        & &  & & & & &
        \\
        %[-0.4cm]
        \ctoprule
        \crowcolor
        Name &
        $U$ & $D$ & $Q_1$ & $Q_5$ & $Q_7$ & $T_1$ & $T_2$ \\
        Irrep &
        $\left(3, 1,\frac{2}{3}\right)$ &
        $\left(3, 1,-\frac{1}{3}\right)$ &
        $\left(3, 2,\frac{1}{6}\right)$ &
        $\left(3, 2,-\frac{5}{6}\right)$ &
        $\left(3, 2,\frac{7}{6}\right)$ &
        $\left(3, 3,-\frac{1}{3}\right)$ &
        $\left(3, 3,\frac{2}{3}\right)$ \\[1.3mm]
        \cbottomrule
      \end{tabular}
    }
\caption{Fermion exits: New vector-like fermions that can couple
linearly to standard model particles. Symbols are  taken from 
\cite{deBlas:2017xtg}. \vspace*{-1cm}}
    \label{t:fermions}
  \end{center}
\vspace{1cm}
\end{table}

\subsection{Tree-level $\psi^2H^2D$ and $\psi^2 H^3$ 
operators\label{sect:h2psi2}}

As discussed above, in this paper we study models that generate
4F operators only at the one-loop level. However, choosing
models which contain one of the exit particles of Tables\,
\ref{t:scalars} and \ref{t:fermions}, some operators from two other
classes, $\psi^2H^2D$ and $\psi^2 H^3$, will be generated at
tree-level. Table~\ref{tab:2fnh_operators} shows all operators
involving two fermions and two Higgs bosons and one derivative or three
Higgs bosons. Anyway, for our models not all of these operators are generated by each
model. Which of these operators is generated depends on the exit
particle of the model under consideration. 

\begin{table}[]
    \centering
    \begin{tabular}{|l|l|l|}
        \hline
        class & name & structure \\
        \hline \hline
        $\psi^2 \phi^2 D$ & $\mathcal{O}_{\phi l}^{(1)}$ & $\left( \phi^{\dagger} i \overset{\leftrightarrow}{D}_{\mu} \phi\right) \left(\bar{l}_L \gamma^{\mu} l_L\right)$ \\
        & $\mathcal{O}_{\phi l}^{(3)}$ & $\left( \phi^{\dagger} i \overset{\leftrightarrow}{D}_{\mu}^a \phi\right) \left(\bar{l}_L \gamma^{\mu} \sigma^a l_L\right)$ \\
        \hline
        & $\mathcal{O}_{\phi e}$ & $\left( \phi^{\dagger} i \overset{\leftrightarrow}{D}_{\mu} \phi\right) \left(\bar{e}_R \gamma^{\mu} e_R \right)$ \\
        \hline
        & $\mathcal{O}_{\phi q}^{(1)}$ &  
        $\left( \phi^{\dagger} i \overset{\leftrightarrow}{D}_{\mu} \phi\right) \left(\bar{q}_L \gamma^{\mu} q_L\right)$ \\
        & $\mathcal{O}_{\phi q}^{(3)}$ & $\left( \phi^{\dagger} i \overset{\leftrightarrow}{D}_{\mu}^a \phi\right) \left(\bar{q}_L \gamma^{\mu} \sigma^a q_L\right)$ \\
        \hline
        & $\mathcal{O}_{\phi u}$ & $\left( \phi^{\dagger} i \overset{\leftrightarrow}{D}_{\mu} \phi\right) \left(\bar{u}_R \gamma^{\mu} u_R \right)$ \\
        \hline
        & $\mathcal{O}_{\phi d}$ & $\left( \phi^{\dagger} i \overset{\leftrightarrow}{D}_{\mu} \phi\right) \left(\bar{d}_R \gamma^{\mu} d_R \right)$ \\
        \hline
        & $\mathcal{O}_{\phi u d}$ & $\left( \phi^{\dagger} i \overset{\leftrightarrow}{D}_{\mu} \phi\right) \left(\bar{u}_R \gamma^{\mu} d_R \right)$ \\
        \hline
        \hline
        $\psi^2 \phi^3$ &  $\mathcal{O}_{e\phi}$ & $\left( \phi^{\dagger} \phi\right)^2 \left(\bar{l}_L \phi e_R \right)$ \\
        \hline
        & $\mathcal{O}_{u\phi}$ & $\left( \phi^{\dagger} \phi\right)^2 \left(\bar{q}_L \phi u_R \right)$ \\
        \hline
        & $\mathcal{O}_{d\phi}$ & $\left( \phi^{\dagger} \phi\right)^2 \left(\bar{q}_L \phi d_R \right)$ \\
        \hline
    \end{tabular}
    \caption{Fermion-Higgs operators at dimension-6 in the Warsaw basis.}
    \label{tab:2fnh_operators}
\end{table}

%%%%%%%%%%%%%%%%%%%%%%%%%%%%%%%%%%
%%%%%%%%%%%%%%%%%%%%%%%%%%%%%%%%%%
\section{Model candidates and  their matching to 4F operators}
\label{sec:model_candidates}

In this section we will count the number of UV completions leading to
loop-suppressed 4F operators, describe their matter content and
discuss some common aspects they share.

\subsection{Counting UV completions}

The Warsaw basis counts 25 baryon number conserving 4F operator
structures at dimension-6, cf. Table \ref{tab:4-fermion_operators},
suppressing flavour indices. Note that while the model diagrams can be
understood as having open flavour indices, in practice we will be
interested only in flavour diagonal operators. The reason for this
restriction is that for flavour-violating operators experimental
constraints are usually much stronger than for flavour conserving
ones, in particular if the operators contain leptons
\cite{Crivellin:2017rmk}.

In the following counting, for simplicity, we will not distinguish between singlet and
non-singlet operators with the same fermion content, for example we
simply write $\mathcal{O}_{lq}$ for $\mathcal{O}_{lq}^{(1)}$ and
$\mathcal{O}_{lq}^{(3)}$. Thus, the basis simplifies to 18
B-conserving 4F operators which can be classified into 3
categories:

\begin{enumerate}
    \item 3 lepton-specific operators (in the following referred to as
      LL): $\mathcal{O}_{ll}$, $\mathcal{O}_{le}$ and
      $\mathcal{O}_{ee}$
    \item 7 quark-specific operators (QQ): $\mathcal{O}_{qq}$,
      $\mathcal{O}_{quqd}$, $\mathcal{O}_{qu}$, $\mathcal{O}_{qd}$,
      $\mathcal{O}_{uu}$, $\mathcal{O}_{ud}$, $\mathcal{O}_{dd}$
    \item 8 mixed lepton-quark operators (LQ): $\mathcal{O}_{lq}$,
      $\mathcal{O}_{lu}$, $\mathcal{O}_{ld}$, $\mathcal{O}_{lequ}$,
      $\mathcal{O}_{leqd}$, $\mathcal{O}_{qe}$, $\mathcal{O}_{eu}$,
      $\mathcal{O}_{ed}$
\end{enumerate}

\begin{figure}[t!]
    \centering
    \includegraphics[scale=0.5]{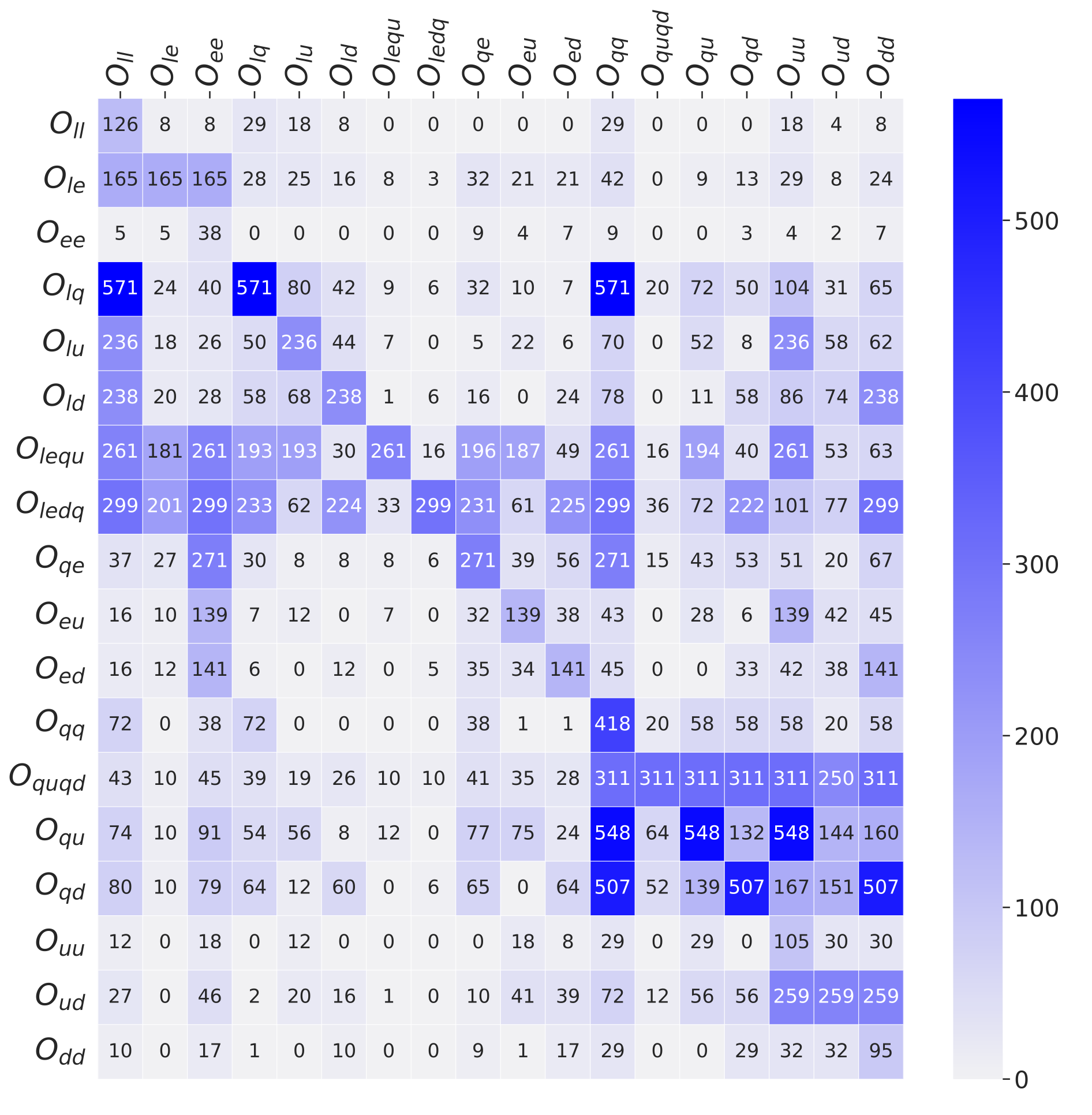}
    \caption{Operator overlap matrix for $maxSU3=8$, $maxSU2=6$ and $maxY = 5$.
      The table counts the number of models for each operator in the
      diagonal, the off-diagonal counts the ``model overlap'', see
      text.}
    \label{fig:confusion_matrix_86}
\end{figure}

The diagrammatic method, discussed in the previous section, has been
implemented into a Mathematica code, which we call the
\textit{ModelGenerator}. Using the \textit{ModelGenerator} we have
found all UV models that open up any of these 18 operator structures
at one-loop and meet the criteria discussed in
section~\ref{sec:UV_model_constraints}.

Moreover, we still have the freedom to define a maximal size of the
representation we want to allow the UV model particles to have under
the SM gauge groups, i.e.\ we can specify maximal values for SU(3)$_C$,
SU(2)$_L$ and the hypercharge Y. We call these $maxSU3$, $maxSU2$ and
$maxY$ in the following. Considering all of the exit particles in Tables
\ref{t:scalars} and \ref{t:fermions}, it is easy to determine that the
maximal possible representation sizes (and numbers) in our models are
$maxSU3 = 8$, $maxSU2 = 6$, $maxY = 5$.

Once we found the models for all the 18 operators, and before we study
some models in detail, it is interesting to cast a glance on the
interplay between the operators and to analyse which models found for
one operator can also open up one of the other operators at one-loop
level. Depending on the structure of the operator, a model can contain
between two and four different BSM particles. So picking one model and
rearranging the particles in the loop or picking a subset of the
particles and repeating some of them can lead to a box-diagram for
another 4-fermion operator.

\begin{figure}[t!]
    \centering
    \includegraphics[scale=0.5]{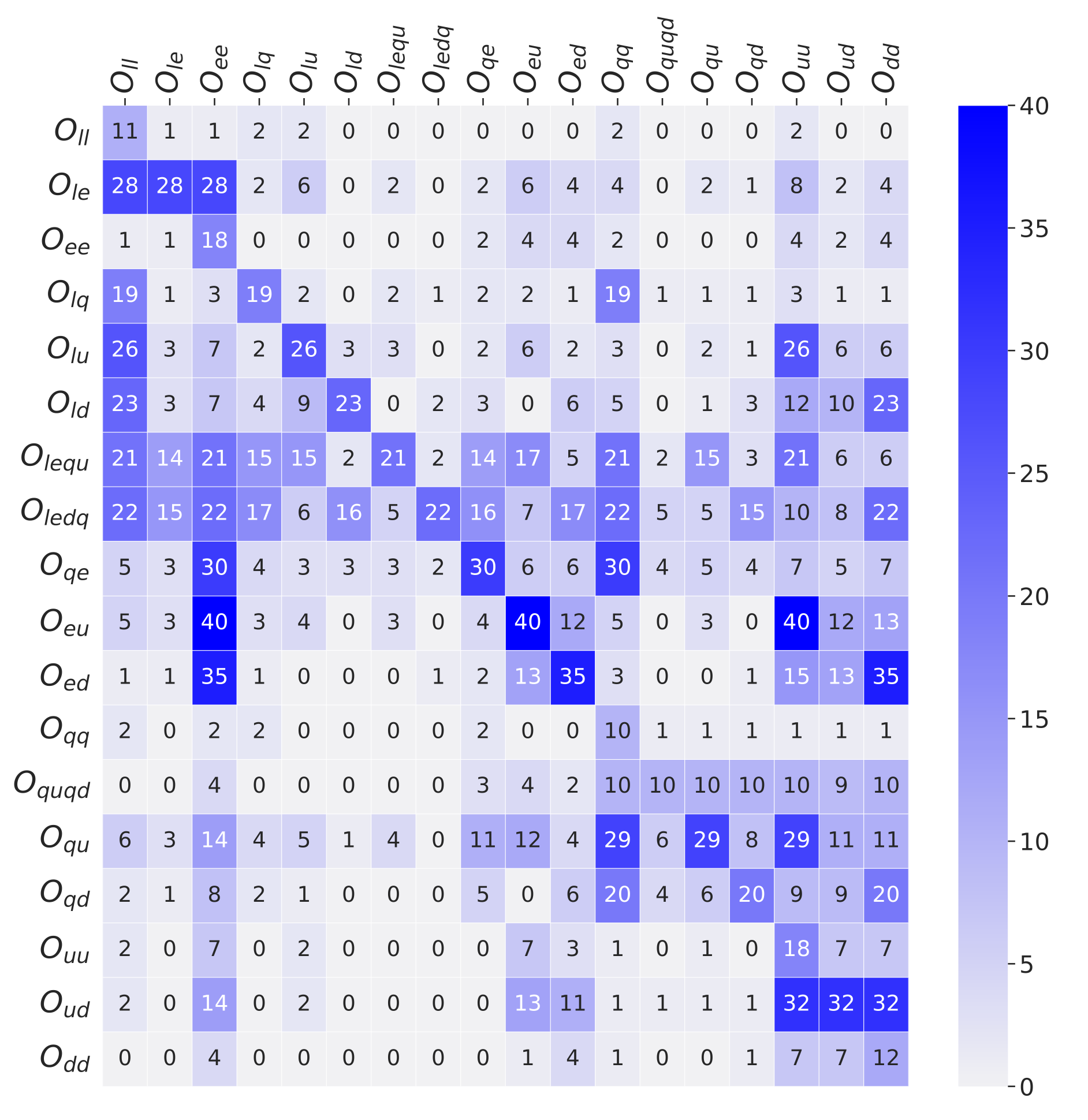}
    \caption{Operator overlap for $maxSU3=3$, $maxSU2=2$ and $maxY = 4$.}
    \label{fig:confusion_matrix_32}
\end{figure}

In Fig.~\ref{fig:confusion_matrix_86} we list the number of models
found for one operator that can open up any of the other 17 4-fermions
operators for the maximal possible representation $maxSU3 =8$,
$maxSU2 = 6$ and $maxY = 5$. To emphasise the large hierarchy between the entries in
the matrix, we underlie the table with a heat-map, with darker colours
representing larger numbers.

The table is to be read in the following way: The diagonal entries
list the total number of models for this operator. Every row belongs
to one operator, say $\mathcal{O}_X$. Then, every entry in this row
lists how many of the models for operator $\mathcal{O}_X$ provide a
contribution to the operator that labels the column.

To give an explicit example: the third row states that we found 38
models for the operator $\mathcal{O}_{ee}$ out of which 5 open up
$\mathcal{O}_{ll}$, 5 $\mathcal{O}_{le}$, 9 $\mathcal{O}_{qe}$, 4
$\mathcal{O}_{eu}$, 7 $\mathcal{O}_{ed}$, 9 $\mathcal{O}_{qq}$, 3
$\mathcal{O}_{qd}$, 4 $\mathcal{O}_{uu}$, 2 $\mathcal{O}_{ud}$ and 7
$\mathcal{O}_{dd}$. All the other entries in this row being zero means
that none of the models for $\mathcal{O}_{ee}$ can be recasted in a
box that contributes to $\mathcal{O}_{lq}$, $\mathcal{O}_{lu}$,
$\mathcal{O}_{ld}$, $\mathcal{O}_{lequ}$, $\mathcal{O}_{ledq}$,
$\mathcal{O}_{quqd}$ or $\mathcal{O}_{qu}$.

As one can see from Fig.~\ref{fig:confusion_matrix_86}, the number of
possible models is still large (571 models for $\mathcal{O}_{lq}$, for
example). Since it is nearly impossible to study the phenomenology for
all of these, it is useful to reduce the number of models by limiting
the maximal allowed representations. The exact choice for $maxSU3$,
$maxSU2$ and $maxY$ is of course somewhat arbitrary. However, since
this choice is a simple post-selection on a larger existing model
list, such cuts can be done straight-forwardly for any choice of
$maxSU3$, $maxSU2$ and $maxY$ one may wish to consider.  In
Fig.~\ref{fig:confusion_matrix_32} we show the overlap matrix for
the concrete choice of $maxSU3=3$, $maxSU2=2$ and $maxY = 4$. The matrix is
obviously much sparser. For example, for $\mathcal{O}_{lq}$ only 19
models remain. Most interesting, however, is that in this table many
more zeroes appear. Thus, it would become much easier to gain
information on the underlying UV model if some non-zero coefficients
for the 4F operators had been established experimentally.

In the following subsections we will discuss some benchmark models for
the three different operator classes, LL, LQ and QQ.

\subsection{Patterns in SMEFT 4F operators}

The inspection of Fig.~\ref{fig:confusion_matrix_86}~and~\ref{fig:confusion_matrix_32} shows that there is still a large number of
possible models even after imposing all the simplifying assumptions we
described in subsection~\ref{sec:UV_model_constraints}. However, the model
overlap matrix becomes quite sparse, especially when focusing on lower
representations of the SM gauge groups,
Fig.~\ref{fig:confusion_matrix_32}.

Also, it is clear that the tables are not symmetric due to the
structure of the different operators. This means that if we have $n$
models for operator $\mathcal{O}_2$ that open up operator
$\mathcal{O}_1$, it does not necessarily imply that any of the models
for operator $\mathcal{O}_1$ also opens up $\mathcal{O}_2$. For
example, consider an operator $\mathcal{O}_1$ generated by some
model with only two different particles, $F_1$ and $S_1$,
such that the box opening reads as $\mathcal{M}_1 = \{F_1, S_1, F_1,
S_1\}$.  Then, for any viable combination of particles $F_2$ and
$S_2$, the model $\mathcal{M}_2 = \{F_1, S_1, F_2, S_2\}$ will generate
operator $\mathcal{O}_2$, as well as $\mathcal{O}_1$ via the same
model $\mathcal{M}_1$, which is a subset of $\mathcal{M}_2$. But, as
long as no opening of $\mathcal{O}_1$ features the particles $F_2$ and
$S_2$, no model for $\mathcal{O}_1$ will open up $\mathcal{O}_2$.

\begin{figure}
    \begin{subfigure}[b]{0.48\textwidth}
        \centering
        \includegraphics[width=0.8\textwidth]{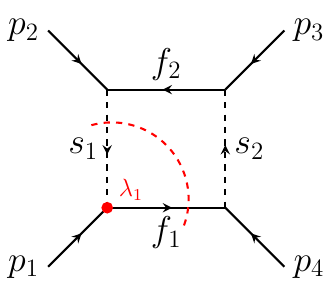}
        \subcaption{Generic 4F operator with up to four different external legs.}
	\end{subfigure}
	\hfill
    \begin{subfigure}[b]{0.48\textwidth}
        \centering
        \includegraphics[width=0.8\textwidth]{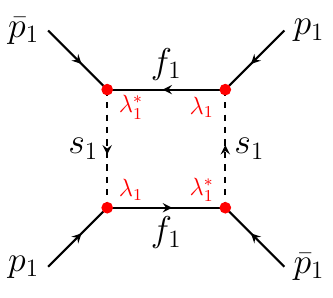}
        \subcaption{Generic 4F operator with all four legs being the same SM particle.}
	\end{subfigure}
    \caption{Schematic example how to obtain models for \textit{more symmetric} 4F operators from models for \textit{less symmetric} ones. A model that generates an operator of the form $\mathcal{O}_{p_1 p_2 p_3 p_4}$ will automatically produce also the operators $\mathcal{O}_{p_1 p_1}$, $\mathcal{O}_{p_2 p_2}$, $\mathcal{O}_{p_3 p_3}$ and $\mathcal{O}_{p_4 p_4}$. See text for details.}
    \label{fig:OLQvsOLL}
\end{figure}

To see an explicit example, consider $\mathcal{O}_{le}$ and $\mathcal{O}_{ll}$:  For $maxSU3=3$, $maxSU2=2$ and $maxY=4$ all
28 models for $\mathcal{O}_{le}$ also open up
$\mathcal{O}_{ll}$, but only one out of 11 models for
$\mathcal{O}_{ll}$ opens $\mathcal{O}_{le}$. As mentioned before, this
is the case because all model-diagrams for $\mathcal{O}_{le}$ generate also $\mathcal{O}_{ll}$. A schematic example of how this works is shown in figure~\ref{fig:OLQvsOLL}: Identifying $p_1$ with the lepton doublet $l$ and being $(p_2, p_3, p_4)$ any permutation of $(\bar{l}, e_R, \bar{e}_R)$, the left side presents a box diagram for $\mathcal{O}_{le}$. Introducing the coupling $\lambda_1$ between $p_1$, $s_1$ and $\bar{f}_1$, we note that cutting the lower left corner and stitching it together with itself and two copies of its complex conjugate leads to a diagram for $\mathcal{O}_{ll}$, as it is shown on the right side of Fig.~\ref{fig:OLQvsOLL}.

On the other hand, finding a model that generates $\mathcal{O}_{le}$ from the list of models that opens $\mathcal{O}_{ll}$ is not straightforward. Due to the symmetry of the external particles of $\mathcal{O}_{ll}$, in general less fields are needed to generate this operator and, thus, one would have to add more particles to obtain a model for $\mathcal{O}_{le}$. The one model for $\mathcal{O}_{ll}$ that opens up $\mathcal{O}_{le}$ features only the particles $S$ and $\Delta_1$: $(S, \Delta_1, S, \bar{\Delta}_1)$. This is a rather particular case where the SM Higgs, along with $S$ and $\Delta_1$, participates in the box diagram that generates $\mathcal{O}_{le}$.

This analysis allows us to start identifying common patterns among
models. First, many of these scenarios exclusively produce LL or QQ
types of operators. Second, all of the models which produce LQ
operators also produce LL and QQ operators. The argument for this is
very similar to the one provided above for $\mathcal{O}_{le}$ and
$\mathcal{O}_{ll}$: Consider an operator $\mathcal{O}_{p_1 p_2 p_3 p_4}$ of the class LQ, i.e. two p's are quarks and two of them are leptons. Then, 
every model for $\mathcal{O}_{p_1 p_2 p_3 p_4}$ will produce
$\mathcal{O}_{p_1 p_1}$, $\mathcal{O}_{p_2 p_2}$, $\mathcal{O}_{p_3 p_3}$ and $\mathcal{O}_{p_4 p_4}$ by taking a subset of the two internal particles of the original model which are adjacent to the respective $p_i$, exactly as it has been explained for $\mathcal{O}_{le}$ and $\mathcal{O}_{ll}$, see Fig.~\ref{fig:OLQvsOLL} again.

This means
that, for instance, all models for $\mathcal{O}_{eq}$ will produce both
$\mathcal{O}_{ee}$ and $\mathcal{O}_{qq}$, while for example the models for
$\mathcal{O}_{ledq}$ will contribute to all of the four operators
$\mathcal{O}_{ll}$, $\mathcal{O}_{ee}$, $\mathcal{O}_{qq}$ and
$\mathcal{O}_{dd}$.

Based on these observations, we can classify three types of scenarios:

\begin{itemize}
    \item {\bf Lepton-specific scenarios:} exclusively producing 4F
      involving leptons.
    \item {\bf Quark-specific scenarios:} constrained to affecting
      hadronic observables.
    \item {\bf Generic or hybrid scenarios:} which contribute to the
      three types of 4F operators.
\end{itemize}

Within each of these categories, we find models with different matter
content. A further simplifying criterion one imposes is {\it minimality},
i.e. to identify which specific models contain the least number of new
particles and hence of new parameters. In subsection \ref{sec:qspec} 
we will discuss the quark-specific models in a bit more detail.

\subsection{New particles running in loop-induced 4F operators} 
\label{sec:complex}

We now examine the actual matter content in these models. As expected,
new types of particles will appear in the UV completions and will be
shared among models.
In particular, if we restrict ourselves to a maximum representation of
triplets for SU(3)$_c$ and doublets for SU(2)$_L$, the list of new
quantum numbers which show up in the UV completions is not long. We
show those new particle candidates in Tables~\ref{t:new_fermions} and
\ref{t:new_scalars}, which we label following the pattern used in the
Granada convention~\cite{deBlas:2017xtg}.

\begin{table}[h!]
  \begin{center}
    {\small
      \begin{tabular}{lcccccccc}
        \ctoprule
        \crowcolor
        Name &
     $Q_{11}$ & $Q_{13}$  & $Q_{17}$ & $X_4$ & $X_5$ & $X_7$ & $\Delta_5$ & $N_2$  \\
        Irrep &
        $\left(3, 2, -\frac{11}{6} \right)$ &
        $\left(3, 2, \frac{13}{6} \right)$ &
        $\left(3, 2, -\frac{17}{6} \right)$ &
        $\left(3, 1, -\frac{4}{3} \right)$ &
        $\left(3, 1, \frac{5}{3} \right)$ &
        $\left(3, 1, -\frac{7}{3} \right)$ &
        $\left(1, 2, \frac{5}{2} \right)$ &
        $\left(1, 1, 2 \right)$
        \\[1.3mm]
       \cbottomrule
      \end{tabular}
    }
\caption{New vector-like fermions.}
    \label{t:new_fermions}
  \end{center}
\end{table}

New fermions $Q_a$ and $X_a$, similar to quarks $Q_L$ and $u_R$ and
$d_R$, but with unusual hypercharges, appear in many realisations. New
colour singlets $\Delta_5$ and $N_2$, similar to the leptons in the
SM, also appear with non-standard hypercharges as well.

\begin{table}[h!]
  \begin{center}
    {\small
      \begin{tabular}{lccccccc}
        \ctoprule
        \crowcolor
        Name &
     $\Pi_5$ & $\Pi_{11}$ & $\Pi_{13}$ & $\omega_5$ & $\phi_3$   \\
        Irrep &
        $\left(3, 2,-\frac{5}{6}\right)$ &
        $\left(3, 2,-\frac{11}{6}\right)$ &
        $\left(3, 2,\frac{13}{6}\right)$ &
        $\left(3, 1,\frac{5}{3}\right)$ &
        $\left(1, 2,\frac{3}{2}\right)$    \\[1.3mm]
       \cbottomrule
      \end{tabular}
    }
\caption{New scalars.}
    \label{t:new_scalars}
  \end{center}
\end{table}

We also find new coloured scalars $\Pi_a$ and $\omega_5$, similar to
squarks in supersymmetry, but again with different hypercharge
assignments. Finally, a Higgs-like new particle $\phi_3$ with
hypercharge 3/2 also shows up in some of the UV completions.

In the next section~\ref{sec:pheno}, we will discuss the phenomenology
of these exotic new particles. Here, we only list the possible ``decay
channels'' for these particles, dictated by our model constructions. 
Tables~\ref{tab:decay_scalar_exits} and \ref{tab:decay_fermion_exits}
list the decay channels for the allowed exit particles in the
models. Note that the ``decay channels'' shown in these tables are 
shown before electroweak symmetry breaking. The real, observable 
decays therefore involve Higgses ($h^0$) or electroweak gauge 
bosons ($W^{\pm}$ or $Z^0$).

\begin{table}[h]
    \centering
    \begin{tabular}{|c|c|l|}
    \hline
    name & representation & decays \\
    \hline
    \hline
    $\mathcal{S}$ & $(1,1,0)$ & $S \rightarrow H + H^{\dagger}$ \\
    $\Xi$ & $(1,3,0)$ & 
    $\Xi \rightarrow H + H^{\dagger}$\\
    $\Theta_1$ & $(1,4,\frac{1}{2})$ & $\Theta_1 \rightarrow H + H + H^{\dagger}$ \\
    $\Theta_3$ & $(1,4,\frac{3}{2})$ & $\Theta_3 \rightarrow H + H + H$ \\
    \hline
    \end{tabular}
    \caption{Decay channels for scalar exits.}
    \label{tab:decay_scalar_exits}
\end{table}

\begin{table}[h]
    \centering
    \begin{tabular}{|c|c|l|}
    \hline
    name & representation & decays \\
    \hline
    \hline
    N & $(1,1,0)$ & $N \rightarrow l +H$ \\
    E & $(1,1,-1)$ & $E \rightarrow l +H^{\dagger}$ \\
    $\Delta_1$ & $(1,2,-\frac{1}{2})$ & $\Delta_1 \rightarrow e_R + H$ \\
    $\Delta_3$ & $(1,2,-\frac{3}{2})$ &  $\Delta_3 \rightarrow e_R + H^{\dagger}$ \\
    $\Sigma$ & $(1,3,0)$ & $\Sigma \rightarrow l + H$ \\
    $\Sigma_1$ & $(1,3,-1)$ & $\Sigma_1 \rightarrow l + H^{\dagger}$ \\
    U & $(3,1, \frac{2}{3})$ & $U \rightarrow q + H$ \\
    D & $(3,1, -\frac{1}{3})$ & $D \rightarrow q + H^{\dagger}$ \\
    $Q_1$ & $(3,2,\frac{1}{6})$ & $Q_1 \rightarrow u_R + H^{\dagger},$ \\
    &  & $ \quad Q_1 \rightarrow d_R + H$ \\
    $Q_5$ & $(3,2,-\frac{5}{6})$ & $Q_5 \rightarrow d_R + H^{\dagger}$\\
    $Q_7$ & $(3,2, \frac{7}{6})$ & $Q_7 \rightarrow u_R + H$ \\
    $T_1$ & $(3,3,-\frac{1}{3})$ & $T_1 \rightarrow q + H$ \\
    $T_3$ & $(3,3,\frac{2}{3})$ & $T_3 \rightarrow q + H^{\dagger}$\\
    \hline
    \end{tabular}
    \caption{Decay channels for fermion exits.}
    \label{tab:decay_fermion_exits}
\end{table}

The ``decay channels'' (before EWSB) of the non-exit particles can be
found in Table~\ref{tab:decay_scalars_loop} for scalars and Table~\ref{tab:decay_fermions_loop} for fermions. To determine their decay
patterns we used the couplings that necessarily exist for the box
diagrams of the models. Note that the non-exit particles could, in
principle, be lighter than the exit particle list in this table. In
this case, it should be implicitly understood that the exit particle
is off-shell and the complete decay chain reads, for example: $\Pi_5
\rightarrow E^* + q \rightarrow  q+ l +H^{\dagger}$ and similar for 
all other cases. 

\begin{table}[h]
    \centering
    \begin{tabular}{|c|c|l|}
    \hline
    name & representation & decays  \\
    \hline 
    \hline
    $\Pi_5$ &  $(3,2,-\frac{5}{6})$ & $\Pi_5 \rightarrow E + q$ \\
    & & $\Pi_5 \rightarrow \bar{U} + \bar{q}$ \\
    & & $\Pi_5 \rightarrow \Delta_1 + d_R$ \\
    & & $\Pi_5 \rightarrow \Delta_3 + u_R$ \\
    & & $\Pi_5 \rightarrow \bar{Q}_1 + \bar{u}_R$ \\
    & & $\Pi_5 \rightarrow \bar{Q}_7 + \bar{d}_R$ \\
    \hline
    $\Pi_{11}$ & $(3,2,-\frac{11}{6})$ & $\Pi_{11} \rightarrow \Delta_3 + d_R$ \\
    & & $\Pi_{11} \rightarrow \bar{Q}_7 + \bar{u}_R$\\
    \hline
    $\Pi_{13}$ & $(3,2,\frac{13}{6})$ & $\Pi_{13} \rightarrow \bar{\Delta}_3 + u_R$ \\
    \hline
    $\omega_5$ & $(3,1,\frac{5}{3})$ & $\omega_5 \rightarrow \bar{E} + u_R$ \\
    & & $\omega_5 \rightarrow \bar{\Delta}_3 + q$\\
    \hline
    $\phi_3$ & $(1,2,\frac{3}{2})$ & $\phi_3 \rightarrow \bar{Q}_5 + u_R$\\
    & & $\phi_3 \rightarrow Q_7 + \bar{d}_R$ \\
    \hline
    \end{tabular}
    \caption{Decay channels for non-exit scalar loop particles.}
    \label{tab:decay_scalars_loop}
\end{table}

\begin{table}[h]
    \centering
    \begin{tabular}{|c|c|l|}
    \hline
    name & representation & decays  \\
    \hline 
    \hline
    $Q_{11}$ & $(3,2,-\frac{11}{6})$ & $Q_{11} \rightarrow \phi_3^{\dagger} + d_R$ \\
    & & $Q_{11} \rightarrow \omega_5^{\dagger} + \bar{q}$ \\
    & & $Q_{11} \rightarrow \Pi_{13}^{\dagger} + \bar{d}_R$ \\
    \hline
    $Q_{13}$ & $(3,2, \frac{13}{6})$ & $Q_{13} \rightarrow \phi_3 + u_R$ \\
    & & $Q_{13} \rightarrow \Pi_{11}^{\dagger} + \bar{d}_R$ \\
    \hline
    $Q_{17}$ & $(3,2,-\frac{17}{6})$ & $Q_{17} \rightarrow \Pi_{13}^{\dagger}+ \bar{u}_R$ \\
    \hline
    $X_4$ & $(3,1, -\frac{4}{3})$ & $X_4 \rightarrow \phi_3^{\dagger} + q$ \\
    & & $X_4 \rightarrow \omega_5^{\dagger} + \bar{d}_R$ \\
    \hline
    $X_5$ & $(3,1, \frac{5}{3})$ & $X_5 \rightarrow \phi_3 + q$ \\
    & & $X_5 \rightarrow \Pi_{11}^{\dagger} + \bar{q}$ \\
    \hline
    $X_7$ & $(3,1, -\frac{7}{3})$ & $X_7 \rightarrow \omega_5^{\dagger}+\bar{u}_R$ \\
    & & $X_7 \rightarrow \Pi_{13}^{\dagger}+\bar{q}$ \\
    \hline
    $\Delta_5$ & $(1,2,\frac{5}{2})$ & $\Delta_5 \rightarrow \Pi_{11}^{\dagger} + u_R$ \\
    & & $\Delta_5 \rightarrow \Pi_{13} + \bar{d}_R$ \\
    \hline
    $N_2$ & $(1,1,2)$ & $N_2 \rightarrow  \omega_5 + \bar{d}_R$\\
    & & $N_2 \rightarrow  \Pi_{11}^{\dagger} + q$\\
    & & $N_2 \rightarrow  \Pi_{13} + \bar{q}$\\
    \hline
    \end{tabular}
    \caption{Decay channels for non-exit fermion loop particles.}
    \label{tab:decay_fermions_loop}
\end{table}

\subsection{Minimal scenarios and their explicit matching}\label{sec:EU}

After identifying the rich set of new states with exotic quantum
numbers we find in these scenarios, we move to consider minimal
extensions to the SM, i.e.\ models that include as few non-SM particles
as possible.

For lepton- and quark-specific scenarios we can find extensions with
only one BSM fermion and a SM Higgs boson, whereas for a hybrid model
producing LL, QQ and LQ operators at the same time, it is easy to see that we need two
different BSM fermions.

The operators $\mathcal{O}_{ll}$ and $\mathcal{O}_{qq}$ can be opened
up by models that contain only one BSM fermion, for example: $E =
(1,1,-1)$ for $\mathcal{O}_{ll}$ and $U = (3, 1, \frac{2}{3})$ for
$\mathcal{O}_{qq}$. The box diagrams are then completed by the SM
Higgs boson. Models for the operator $\mathcal{O}_{lq}$ require two
different BSM fermions, e.g. $E$ and $U$. The particle content for
each operator is listed in
Table~\ref{tab:particle_content_minimal_models}.

\begin{table}[h]
    \centering
    \begin{tabular}{|c|c|}
    \hline
         operator &  particles \\
         \hline \hline
         $\mathcal{O}_{ll}$ & H, E \\
         \hline
         $\mathcal{O}_{qq}$ & H, U \\
         \hline 
         $\mathcal{O}_{lq}$ & H, E, U \\
         \hline
    \end{tabular}
    \caption{Particle content for minimal models for
      $\mathcal{O}_{ll}$, $\mathcal{O}_{lq}$ and $\mathcal{O}_{qq}$.}
    \label{tab:particle_content_minimal_models}
\end{table}

Accordingly, we will consider the Lagrangian $\mathcal{L}_{UV} =
\mathcal{L}_{SM} + \mathcal{L}_{NP}$ with $\mathcal{L}_{SM}$, the
unbroken SM Lagrangian and the new particle contribution
\begin{equation}
    \mathcal{L}_{NP} = -\lambda_E \bar{E} L H^{\dagger} 
                       -\lambda_U \bar{U} Q H + {\rm h.c.}  
                       - m_E \bar{E} E - m_U \bar{U} U \, .
\end{equation}
We stress here that all exotic fermions are considered to be 
vector-like, i.e. $\bar{E}$ is a different Weyl state and not 
simply the conjugated state of $E$.

This matter content with the new couplings can be found in different
scenarios considered in the literature~\cite{delAguila:2000rc,delAguila:2008pw}. The particle $E$ is a
right-handed massive lepton which mixes with the SM leptons after
EWSB and, similarly, $U$ is an up-type right-handed quark mixing with
the SM quark. We will assume one generation of each of these states,
but more general scenarios could lead to interesting flavour
signatures.

\begin{figure}[t!]
    \begin{subfigure}[b]{0.5\textwidth}
        \centering
        \includegraphics[width=0.8\textwidth]{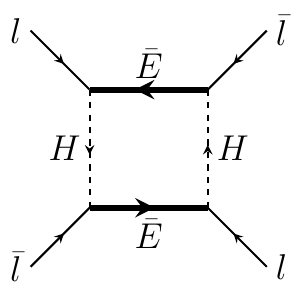}
	\subcaption{$\mathcal{O}_{ll}$}
	\end{subfigure}
	\begin{subfigure}[b]{0.5\textwidth}
        \centering
        \includegraphics[width=0.8\textwidth]{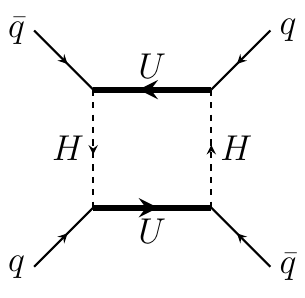}
	\subcaption{$\mathcal{O}_{qq}$}
	\end{subfigure}
	\begin{subfigure}[b]{0.5\textwidth}
        \centering
        \includegraphics[width=0.8\textwidth]{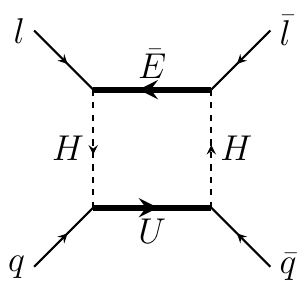}
	\subcaption{$\mathcal{O}_{lq}$}
	\end{subfigure}
    \caption{Minimal models for the operators $\mathcal{O}_{ll}$, $\mathcal{O}_{qq}$ and $\mathcal{O}_{lq}$.}
    \label{fig:minimal_models_LL_QQ_LQ}
\end{figure}

Given this Lagrangian, we can calculate the matching to the SMEFT with
the dimension-6 operators in the Warsaw basis,
cf.\ Table~\ref{tab:4-fermion_operators} by evaluating loop
contributions as shown in Fig.~\ref{fig:minimal_models_LL_QQ_LQ}. 
We performed the exact matching with
\verb|MatchmakerEFT| \cite{Carmona:2021xtq}.

\begin{table}[t!]
    \centering
    \begin{tabular}{|c|c|c|}
    \hline
    Operator & General expression & Equal mass limit \\
    \hline
    \hline
     $c_{ll}$ & $-\frac{1}{8} \frac{1}{16 \pi^2} \frac{\left|\lambda_E\right|^4}{m_E^2}$ & $-\frac{1}{8} \frac{1}{16 \pi^2} \frac{\left|\lambda_E\right|^4}{\Lambda^2}$\\
     \hline
     $c_{lq}^{(1)}$ & $\frac{1}{8} \frac{1}{16 \pi^2} \frac{\left|\lambda_E\right|^2\left|\lambda_U\right|^2 \log\left(\frac{m_E^2}{m_U^2}\right)}{m_E^2-m_U^2}$ & $\frac{1}{8} \frac{1}{16 \pi^2} \frac{\left|\lambda_E\right|^2\left|\lambda_U\right|^2}{\Lambda^2}$ \\
     \hline
     $c_{lq}^{(3)}$ & $-\frac{1}{8} \frac{1}{16 \pi^2} \frac{\left|\lambda_E\right|^2\left|\lambda_U\right|^2 \log\left(\frac{m_E^2}{m_U^2}\right)}{m_E^2-m_U^2}$ & $-\frac{1}{8} \frac{1}{16 \pi^2} \frac{\left|\lambda_E\right|^2\left|\lambda_U\right|^2}{\Lambda^2}$\\
     \hline
     $c_{qq}^{(1)}$ & $-\frac{1}{16} \frac{1}{16 \pi^2} \frac{\left|\lambda_U\right|^4}{m_U^2}$ & $-\frac{1}{16} \frac{1}{16 \pi^2} \frac{\left|\lambda_U\right|^4}{\Lambda^2}$ \\
     \hline
     $c_{qq}^{(3)}$ & $-\frac{1}{16} \frac{1}{16 \pi^2} \frac{\left|\lambda_U\right|^4}{m_U^2}$ & $-\frac{1}{16} \frac{1}{16 \pi^2} \frac{\left|\lambda_U\right|^4}{\Lambda^2}$  \\
     \hline
    \end{tabular}
    \caption{Matching for the simple ``EU'' model in the limit of $\lambda_{SM} \to  0$.}
    \label{tab:matching_UE_model}
\end{table}

The matching of the simple benchmarks to the SMEFT 4F operators in
terms of masses, $m_{E,U}$, and couplings, $\lambda_{E,U}$, to the SM
is shown in Table~\ref{tab:matching_UE_model}, in the limit where all
SM couplings are set to zero. We provide expressions for generic
masses and couplings and in the right-most column the limit for the
couplings and masses being set equal at a scale $\Lambda = m_E =
m_U$. The Wilson coefficients for all other 4F operators are zero in
this limit, consistent with the overlap table.

\begin{figure}[h!]
    \begin{subfigure}[b]{0.5\textwidth}
        \centering
        \includegraphics[width=0.8\textwidth]{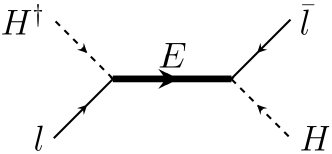}
	\subcaption{$\mathcal{O}_{H l}$}
	\end{subfigure}
	\begin{subfigure}[b]{0.5\textwidth}
        \centering
        \includegraphics[width=0.8\textwidth]{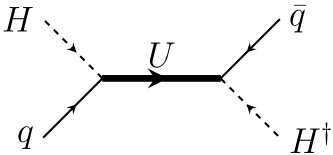}
	\subcaption{$\mathcal{O}_{Hq}$}
	\end{subfigure}
	
	\begin{subfigure}[b]{0.5\textwidth}
        \centering
        \includegraphics[width=0.8\textwidth]{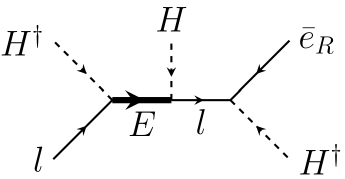}
	\subcaption{$\mathcal{O}_{eH}$}
	\end{subfigure}
	\begin{subfigure}[b]{0.5\textwidth}
        \centering
        \includegraphics[width=0.8\textwidth]{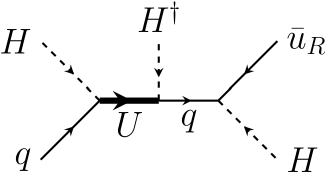}
	\subcaption{$\mathcal{O}_{uH}$}
	\end{subfigure}
    \caption{Diagrams for the operators $\mathcal{O}_{Hl}$,
      $\mathcal{O}_{Hq}$, $\mathcal{O}_{eH}$ and $\mathcal{O}_{uH}$ at
      tree-level. $L$ and $Q$ denote the SM lepton and quark doublet
      respectively.}
    \label{fig:minimal_models_HL_HQ}
\end{figure}

In Table~\ref{tab:matching_UE_model_FH_terms} we also provide the
matching with the tree-level contributions of the model to
Higgs-fermion couplings which would result from computing diagrams as
shown in Fig.~\ref{fig:minimal_models_HL_HQ}.
Note that here we kept the SM Yukawa couplings for completeness. 
Thus, some of these tree-level contributions are Yukawa-suppressed
and would only be relevant for the top quark.

\begin{table}[ht!]
    \centering
    \begin{tabular}{|c|c|}
    \hline
    Operator & General expression \\
    \hline
    \hline
     $c_{\phi l}^{(1)}$ & $-\frac{1}{4} \frac{\left|\lambda_E\right|^2}{m_E^2}$ \\
     \hline
     $c_{\phi l}^{(3)}$ & $-\frac{1}{4} \frac{\left|\lambda_E\right|^2}{m_E^2}$ \\
     \hline
     $c_{\phi q}^{(1)}$ & $\frac{1}{4} \frac{\left|\lambda_U\right|^2}{m_U^2}$ \\
     \hline
     $c_{\phi q}^{(3)}$ & $-\frac{1}{4} \frac{\left|\lambda_U\right|^2}{m_U^2}$  \\
     \hline
    $c_{e \phi}$ & $\frac{1}{2} \frac{\left|\lambda_E\right|^2}{m_E^2} y_e$ \\
    \hline
    $c_{u \phi}$ & $\frac{1}{2} \frac{\left|\lambda_U\right|^2}{m_U^2} y_u$ \\
    \hline
    \end{tabular}
    \caption{Matching for the $\psi H$ terms in the EU model. Since all coefficients depend on only one mass, the equal mass limit $m_E = m_U = \Lambda$ is trivial.}
    \label{tab:matching_UE_model_FH_terms}
\end{table}

\begin{figure}[t]
\centering
\includegraphics[width=0.49\textwidth]{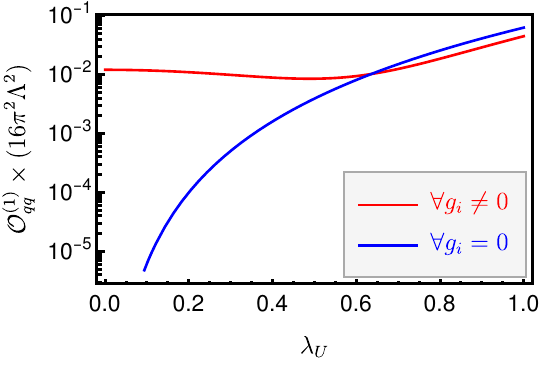}
\includegraphics[width=0.49\textwidth]{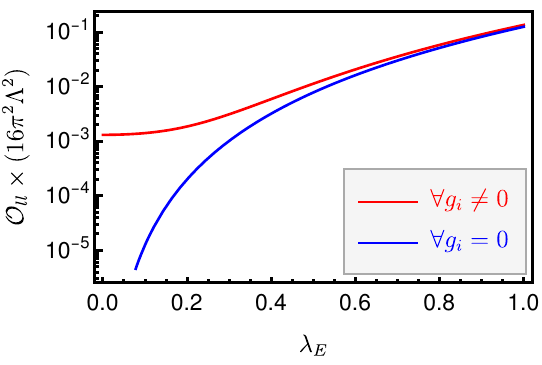}
\caption{The coefficients for the operators ${\cal O}_{qq}^{(1)}$
(left) and ${\cal O}_{ll}$ as a function of the new Yukawa coupling
$\lambda_U$ and $\lambda_E$ respectively. In each case we multiply the coefficient by
a factor $16 \pi^2 \Lambda^2$, where $\Lambda$ is either $M_U$ or
$M_E$, common to all one-loop generated operators. The blue lines
set all SM couplings to zero arbitrarily while the red lines show
the calculation for $g_i$ non-zero. For discussion see text.}
\label{fig:OQQ1OLL}
\end{figure}

The matching discussed above is given in the limit where all SM couplings are neglected. Whether this is a good approximation or
not depends on the actual numerical value of the (unknown) BSM
couplings $\lambda_U$ and $\lambda_E$. In Fig.~\ref{fig:OQQ1OLL}
we show the coefficients of two example operators multiplied
by $16 \pi^2 \Lambda^2$ as a function of $\lambda_U$, $\lambda_E$.
In the limit $\lambda_U$, $\lambda_E \to 1$, the corrections to
the coefficients from the gauge couplings is order 40 \% (10 \%)
in case of ${\cal O}_{qq}^{(1)}$ (${\cal O}_{ll}$). Both larger and
smaller coefficients are possible once the $g_i$'s are taken into
account, the relative sign depending on both the operator and the
model generating it.

In summary, we have discussed some simple scenarios which produce
lepton- and quark-specific 4F operators, and combining both we also
have one example for a minimal hybrid model.
The matching, in the limit of equal masses and couplings and in the
limit of negligible gauge and Yukawa couplings, produces interesting
patterns which we will explore in the next section, devoted to the
interplay between direct searches at the LHC and the low-energy limits
provided by former experiments. The pattern can be summarised as
follows:

\[
 \boxed{\textrm{EU model pattern: }c_{ll} = -c_{lq}^{(1)}= c_{lq}^{(3)}=2 \, c_{qq}^{(1)}= 2\, c_{qq}^{(3)}}, 
 \]
whereas all the other 4F operators are not generated.

\subsection{Non-minimal quark-specific scenarios\label{sec:qspec}}

The models presented in the previous subsection are all minimal in the
sense that they contain as few BSM particles as possible. One
interesting scenario beyond minimality are models that only lead to
4-quark operators and do not produce any 4F operators involving
leptons.  These models, which evade the stronger LL and LQ constraints
from low energies, are of particular interest for hadron collider
searches.

Restricting ourselves to the representations $(maxSU3, maxSU2, maxY) =
(3,2,4)$ we find that only a small number of particles actually
contributes to the models that open up the 4-quark operators. These
models include SM Higgs doublets $H$, the fermionic exits $\Delta_1$,
$\Delta_3$, $U$, $D$, $E$, $Q_1$, $Q_5$, $Q_7$ and the scalar exit
$S$.

The possible decay channels of all the exits were listed in Table
\ref{tab:decay_scalar_exits} for scalars, and in Table
\ref{tab:decay_fermion_exits} for fermions, respectively. Note that by
definition these exits always decay into two SM particles. Moreover,
the models feature 14 non-exit particles which decay into one exit
particle and a SM particle. The description of these new particles can
be found in subsection~\ref{sec:complex} and their decays were listed in
Tables \ref{tab:decay_scalars_loop} and \ref{tab:decay_fermions_loop}.

The list of models which generate 4Q operators are shown in
Appendix~\ref{app:QQ}, in Tables~\ref{tab:quark_specific_models_1} and
\ref{tab:quark_specific_models_2}.
Among these quark-specific
models, models that produce only 4-quark operators and not any
4-lepton or 2-quark-2-lepton operator, are identified with bold
symbols.

Note that more than half of the models which produce 4Q operators are
also quark-specific.  Therefore, from the standpoint of model
building, it is natural to think of UV scenarios which would evade
most low-energy constraints, and for which the LHC would be the most
sensitive probe.

\section{The phenomenological interplay between SMEFT indirect limits 
and direct searches at the LHC}\label{sec:pheno}

So far we have discussed UV completions for 4F operators from a
theoretical perspective. We have produced a list of models leading to
specific 4F operators and discussed assumptions to reduce their
number. For example, when we consider only representations up to
triplets of color and electroweak doublets, we found that the number
of possible scenarios is quite manageable. In this section we discuss
some of the phenomenology that these models would produce.

\subsection{Indirect searches with precise low-energy data}

The first way to search for these scenarios would be through their
imprint at low energies, in terms of 4F operators of the LL, LQ and QQ
types. Constraints on LL~\cite{Falkowski:2015krw} and LQ
operators~\cite{Carpentier:2010ue,deBlas:2013qqa,Falkowski:2017pss}
are particularly strong as they can be probed by high-precision
measurements, particularly at $e^+e^-$ collisions, neutrino scattering
and atomic parity violation experiments.

A global analysis of the constraints derived from low-energy probes in
LL and LQ 4F operators was performed in \cite{Falkowski:2017pss},
where the authors also provided the $\chi^2$ in terms of the Wilson
coefficients. For this work, the EFT effects in 4F are loop-suppressed
by construction. Moreover, in each specific UV completion one finds
relations among the coefficients (patterns) which further restrict the
form of the $\chi^2$ function. For example, in the simple EU model
described in subsection~\ref{sec:EU}, most of the Wilson coefficients are
not generated or they are suppressed by Yukawa or gauge couplings. And
the coefficients that are generated are related to each other in
simple ways when the new couplings and masses are set to be equal,
e.g. the pattern in the EU model looks like
\begin{equation}
    c_{ll} =  -c_{lq}^{(1)}=  c_{lq}^{(3)}=2 \, c_{qq}^{(1)}= 2\, c_{qq}^{(3)} \ .
\end{equation}

This type of behaviour is expected from any UV completion we have
considered here, as in each loop there would be 2-4 new particles and
they would only generate a subset of Wilson coefficients. Patterns
like the one shown above tend to produce stronger constraints on the
Wilson coefficients, as one coefficient can be probed in more than one
type of experiment.

In particular for the EU model, the form of the $\chi^2$ is simple and
depends on one variable
\begin{equation}
    \chi^2(\bar c_{ll})=   26.8+  198.4 \,  \bar  c_{ll} 
                    + 1.42 \times 10^6 \, \bar c_{ll}^2  \ ,
\end{equation}
where we have defined $\bar c_{ll} = v^2
\, c_{ll} $, with $v=$ 246  GeV~\cite{Falkowski:2017pss}.

Minimising $\chi^2-\chi_{min}^2$, we obtain a 2$\sigma$ limit in the
 parameter $\bar c_{ll} \in [-1.7,1.6]\times 10^{-3}$, which translates into 
\begin{equation}\label{consEU}
c^{EU}_{ll} \in [-2.9,2.7] \times 10^{-2}\, \textrm{ TeV}^{-2}   \ .
\end{equation}

\subsection{Direct high-energy probes}

The EFT can also be probed at high energies, exploiting the specific
non-resonant kinematics they produce.  In the context of the LL and LQ
operators, this question was explored in
\cite{Falkowski:2017pss}. Similarly, for individual QQ operators
the best high-energy probes are dijet searches at the
LHC~\cite{Domenech:2012ai}.

In this work, though, we focus on models which induce loop-suppressed
4F operators and render much weaker indirect limits on the UV parameter
space. Within these scenarios, one can explore the interplay between
indirect precision searches and direct LHC resonant searches.  With
light resonances, the extension of the EFT approach from precision
low-energy to the LHC would break down, such as the 4F analyses in
Refs.~\cite{Falkowski:2017pss, Domenech:2012ai}. 

From all the new resonances, one should expect those with colour charge
to be produced more copiously. Among the exit particles, one finds
several vector-like fermion exits triplets of color, $U$, $D$,
$Q_{1,5,7}$ and $T_{1,2}$, see Table~\ref{t:fermions}. Inspecting the
decay Tables~\ref{tab:decay_fermion_exits} and evaluating them after
electroweak symmetry breaking, one finds that the main LHC search for
the UV completions would be diboson production with two energetic
jets, namely
\begin{eqnarray}
    p \, p &\to& h \, h + 2 \, j \nonumber \\
    & & W^{\pm} \, (Z,\gamma) + 2 \, j \nonumber \\
    && W^{+} \, W^{-} + 2 \, j \nonumber \\
     && (Z,\gamma) \, (Z,\gamma) + 2 \, j \ ,
\end{eqnarray}
where one boson and one jet would reconstruct the resonance. 

There are few searches in these final states, particularly focused on
vector-boson fusion (VBF). For example, the di-Higgs VBF production
analysis in Ref~\cite{ATLAS:2020jgy} would capture some of this signal
but with a largely reduced sensitivity due to forward-jet cuts and
veto on centrality. Another possibility would be to explore the
sensitivity of diboson channels and focus on the kinematic region with
two or more hard jets, e.g. reinterpreting the combination of di-Higgs
searches in \cite{ATLAS:2019qdc}.

A copious di-Higgs production is particularly attractive, as it is a
priority for LHC searches but the SM rates are very low. On the other
hand, the best limits are likely to come from well-understood,
high-statistics channels with $W^+ W^-$. In particular, searches of
two W-bosons with more than one hard jet should be particularly
sensitive.

In \cite{ATLAS:2021jgw}, the channel with $e^\pm \nu \mu\mp \nu$
with at least one jet of $p_T> 30$ GeV was explored with the Run2
dataset of 139 fb$^{-1}$. Using the HEPDATA source on the distribution
in jets and focusing on the bin with two jets, the agreement between
prediction and data is at the level of 10\%. This leads to a limit
after cuts on the production cross-section of new particles of around
60-70 fb.  Note that this limit is very rough and sub-optimal, as a
dedicated search for pair production of resonances decaying into a
$W$-boson and a jet would be more sensitive.  We will nevertheless use
this limit based on an existing search to illustrate the interplay of
direct and indirect searches in the next subsection.

We also find several non-exit new coloured particles, see
Tables~\ref{t:new_fermions} and \ref{t:new_scalars} which would be
double-produced via their strong coupling and would lead to even
richer final states, namely states with two bosons and 3-4 additional
objects,
\begin{eqnarray}
     p \, p &\to& \textrm{Diboson }+ 4 \, j  \nonumber \\
    && \textrm{Diboson }+ 2 \, j + 2\, \ell \nonumber \\
    && \textrm{Diboson }+ 2 \, j + \ell^{\pm} + \textrm{ MET} \nonumber \\
    && \textrm{Diboson }+ 2 \, j +  \textrm{ MET}.
\end{eqnarray}

\subsection{The interplay between low-energy and LHC resonant probes}

We finish this section by exploring an example of the complementarity
between indirect and direct probes in the loop-suppressed 4F
scenarios. We choose for simplicity the minimal EU model described in
subsection~\ref{sec:EU}. For simplicity, we will also assume that all new
couplings and masses are equal, which leads to simple matching
expressions, cf.\ the right-most column in Table~\ref{tab:matching_UE_model}.

This model is constrained by low-energies as shown in Eq.~\eqref{consEU}, 
which translates into a rather weak 2$\sigma$ constraint on the combination of
model parameters,  $m_{EU} > 0.17 \,  |\lambda_{EU}|^2$ TeV.

As discussed in the previous section, pairs of coloured resonances
would be strongly produced, and decay to a two-boson and two-jet final
state. We find that the current most-sensitive channel is the SM
$W^+W^-$ measurement at the LHC in the two-jet channel, where we find
an approximate 2$\sigma$ cross-section limit of 65 fb. Assuming order one branching ratio to this final state and efficiency to the
experimental cuts also of order one, this limit in cross-section leads
to a mass limit of the order of 650 GeV.

\begin{figure}[t!]
\begin{center}
\includegraphics[width=4in]{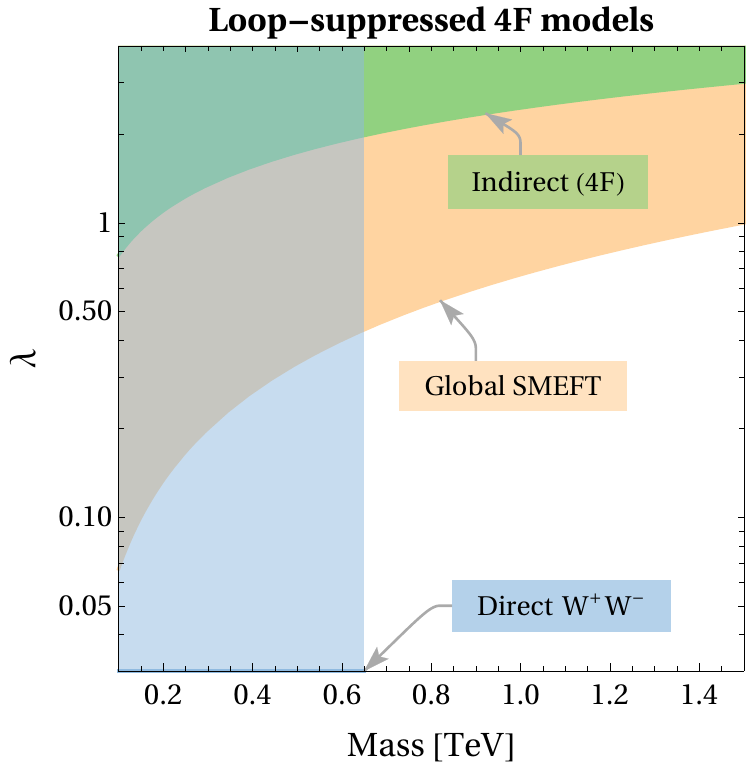}
\caption{Interplay between indirect low-energy probes, SMEFT global fits, and resonant LHC
  searches for the EU benchmark. }
\label{phenointerplay}
\end{center}
\end{figure}

Besides providing direct limits on resonant production, the LHC  provides indirect information on new physics. In particular, global SMEFT fits using low-energy data and LHC SM precision measurements place bounds on contributions to Higgs-fermion dimension-six couplings, such as ${\cal O}_{\phi l}^{(1,3)}$ and ${\cal O}_{\phi q}^{(1,3)}$. Following the global analysis in \cite{Ellis:2018gqa} we know the bounds on SMEFT operators after marginalising over possible SMEFT contributions. In particular, we find that the most stringent limit in the set of operators  ${\cal O}_{\phi l}^{(1,3)}$ and ${\cal O}_{\phi q}^{(1,3)}$  is a 2$\sigma$ bound $c_{\phi q}^{(3)} \in  [-0.11, +0.012]$ TeV$^{-2}$, with the other three operators constrained at a similar, but weaker, level.

In the EU model, the translation between operator coefficients and model parameters is shown in Table~\ref{tab:matching_UE_model_FH_terms}, leading to $m_{EU} > |\lambda_{EU}| 1.5$ TeV, which is, as expected, much stronger than the loop-suppressed 4F limits.

The combination of the direct and indirect limits is shown in
Fig.~\ref{phenointerplay}. From this plot, one observes that the LHC direct probes
provide a better handle on scenarios with relatively small coupling
$\lambda\sim $ 0.5 than the SMEFT indirect probes. 

We are comparing in the same mass-coupling plot indirect and direct probes, but please note that  SMEFT global fits
such as \cite{Ellis:2020unq,Ellis:2018gqa} could be invalid if the observables used to set the limits were probing scales close or below $m_{EU}$.

It is  also worth noticing that 
indirect probes using SM measurements at the LHC and the LHC direct
searches may not have the same scaling with  luminosity.  For example, one should expect that precision on diboson channels with higher jet multiplicity will improve with Run3
all the way to the HL-LHC phase, whereas the SMEFT global fits may soon  saturate for the fermion-Higgs operators in Table~\ref{tab:2fnh_operators}, see \cite{Ellis:2018gqa} for more details.

Although Fig.~\ref{phenointerplay} shows an interpretation in terms of
a simple model, and we have done a rough analysis of the LHC reach, it
illustrates the point that UV models with loop-suppressed 4F operators
are interesting for the LHC physics case. As we have seen, these UV
models contain particles with QCD coupling, unusual quantum
numbers, and rich final states which motivate new types of searches.

\section{Conclusions}\label{sec:concls}

In this paper we have focused on identifying UV scenarios leading to
loop-suppressed four-fermion (4F) operators. Our motivation was to
identify classes of UV models which could be discovered at the LHC,
despite contributing to precise low-energy measurements.

To find these models we classified the topologies leading to
loop-suppressed 4F, a task we performed using a diagrammatic
approach. Although the method is general, to present our results we
made various assumptions including:
\begin{itemize}
    \item We considered UV models which do not contain a stable
      particle, a condition we ensure by imposing that one of the
      particles in the loop is an {\it exit} particle, namely it can
      decay to two SM particles. This assumption would be easily
      lifted to consider UV models with Dark Matter candidates.
    \item We focused on dimension-six operators, although our method
      could be used to build topologies with higher-order
      operators. This could be particularly interesting to identify UV
      completions with suppressed dimension-six operators, leading to
      a more dramatic energy behaviour of amplitudes, see
      e.g.~\cite{Degrande:2013kka} where the dimension-eight
      modifications of the gauge couplings are considered.
    \item In the phenomenological analysis we neglected SM couplings,
      although those contributions could be important when the new
      physics couplings become small, see subsection~\ref{sec:EU} for a
      brief discussion.
\end{itemize}

To connect with phenomenology, we classified the 4F operators as
lepton- or quark-specific operators (those involving only leptons or
quarks), and mixed operators. Low-energy limits for operators
involving leptons are particularly strong, whereas quark-specific
operators are especially interesting for the LHC.

We found that the typical LHC signatures of scenarios with
loop-suppressed 4F operators are quite distinct, namely the production
of an even number of bosons ($h$, $W^\pm$, $Z$ and $\gamma$) and
several jets and leptons. Those are new types of searches which are
now motivated by the complementarity between SMEFT limits from low
energies and resonances at the LHC.

To exemplify this complementarity, we provided a simple benchmark
producing mixed operators and estimated the reach from the SM LHC
measurement of $W^+W^-$ in the 2 jet bin. We found that the LHC
already has a good reach, exploring areas in parameter space the low
energy measurements cannot access.  This naive example, based on
recasting a SM measurement, suggests that dedicated searches in the
multi-boson plus leptons and/or jets could reach further than our
estimated 650 GeV.

\section*{Acknowledgements}

We would like to acknowledge discussions with Martin Gonzalez-Alonso,
Arsenii Titov and Avelino Vicente. We would also like to thank Jose
Santiago for help with the tool MatchMakerEFT~\cite{Carmona:2021xtq}.
FE is supported by the Generalitat Valenciana with the grant
GRISOLIAP/2020/145. The research of VS is supported by the Generalitat
Valenciana PROMETEO/2021/083 and the Ministerio de Ciencia e
Innovacion PID2020-113644GB-I00. M.H. acknowledges support by grants
PID2020-113775GB-I00 (AEI/10.13039/ 501100011033) and
CIPROM/2021/054 (Generalitat Valenciana).  R.C. is supported by the
Alexander von Humboldt Foundation Fellowship.

\appendix

\section{Quark specific models}~\label{App:quark_models}\label{app:QQ}

We list here the models that produce 4-quark operators for $maxSU3=3$, $maxSU2=2$ and $maxY = 4$ at one-loop level. In bold-face we denote those models which are
quark-specific, namely those that do not produce any LL and LQ 4F
operators. The tables actually list the particles internal to the 
corresponding box diagrams, therefore also $H$ is listed, despite 
being part of the SM particle content.

\begin{table}[ht]
    \centering
    \begin{tabular}{|c|cccc|}
    \hline
    operator & & & & \\
    \hline
    \hline
    $\mathcal{O}_{qq}$   
    & $\mathbf{H}$ & $\mathbf{U}$ & $\mathbf{H}$ & $\mathbf{\bar{D}}$ \\
    & $\mathbf{U}$ & $\mathbf{\Pi_5^{\dagger}}$ & $\mathbf{U}$ & $\mathbf{H}$ \\
    & $\mathbf{U}$ & $\mathbf{H}$ & $\mathbf{U}$ & $\mathbf{H}$ \\
    & $\mathbf{Q_1}$ & $\mathbf{S}$ & $\mathbf{Q_1}$ & $\mathbf{S}$ \\
    & $\mathbf{D}$ & $\mathbf{H^{\dagger}}$ & $\mathbf{D}$ & $\mathbf{H^{\dagger}}$ \\
    & $\mathbf{\bar{U}}$ & $\mathbf{\Pi_5}$ & $\mathbf{\bar{U}}$ & $\mathbf{\Pi_5}$ \\
    & $\omega_5$ & $\bar{Q}_{11}$ & $\omega_5$ & $\bar{\Delta}_3$ \\
    & $\Pi_5^{\dagger}$ & $\bar{E}$ & $\Pi_5^{\dagger}$ & $U$\\
    & $\bar{E}$ & $\Pi_5^{\dagger}$ & $\bar{E}$ & $\Pi_5^{\dagger}$ \\
    & $\Delta_3$ & $\omega_5^{\dagger}$ & $\Delta_3$ & $\omega_5^{\dagger}$ \\
    \hline
    $\mathcal{O}_{quqd}$ 
    &$\mathbf{\bar{D}}$ & $\mathbf{H}$ & $\mathbf{U}$ & $\mathbf{S}$ \\
    & $\mathbf{\bar{Q}_1}$ & $\mathbf{S}$ & $\mathbf{Q_1}$ & $\mathbf{H^{\dagger}}$ \\
    & $\mathbf{Q_7}$ & $\mathbf{H}$ & $\mathbf{U}$ & $\mathbf{\Pi_5^{\dagger}}$ \\
    & $\mathbf{\bar{D}}$ & $\mathbf{H}$ & $\mathbf{\bar{Q}_1}$ & $\mathbf{S}$ \\
    & $\mathbf{\bar{Q}_1}$ & $\mathbf{S}$ & $\mathbf{\bar{U}}$ & $\mathbf{H^{\dagger}}$ \\
    & $\mathbf{\bar{Q}_7}$ & $\mathbf{\Pi_{11}}$ & $\mathbf{\bar{X}_5}$ & $\mathbf{\phi_3^{\dagger}}$ \\
    & $\bar{Q}_1$ & $\Pi_5$ & $\bar{U}$ & $H^{\dagger}$ \\
    & $\Delta_3$ & $\Pi_{13}^{\dagger}$ & $\bar{N}_2$ & $\Pi_{11}$ \\
    & $N_2$ & $\Pi_{13}$ & $\bar{\Delta}_3$ & $\omega_5$ \\
    & $Q_{11}$ & $\omega_5^{\dagger}$ & $\Delta_3$ & $\Pi_{13}^{\dagger}$ \\
    \hline
    $\mathcal{O}_{qd}$ 
    & $\mathbf{\phi_3}$ & $\mathbf{X_5}$ & $\mathbf{\phi_3}$ & $\mathbf{Q_7}$ \\
    & $\mathbf{H}$ & $\mathbf{U}$ & $\mathbf{H}$ & $\mathbf{Q_1}$ \\
    & $\mathbf{\bar{D}}$ & $\mathbf{H}$ & $\mathbf{Q_1}$ & $\mathbf{S}$ \\
    & $\mathbf{\bar{Q}_5}$ & $\mathbf{H}$ & $\mathbf{U}$ & $\mathbf{H}$ \\
    & $\mathbf{\bar{D}}$ & $\mathbf{S}$ & $\mathbf{Q_1}$ & $\mathbf{S}$ \\
    & $\mathbf{H^{\dagger}}$ & $\mathbf{D}$ & $\mathbf{H^{\dagger}}$ & $\mathbf{Q_5}$ \\
    & $\mathbf{\Pi_5}$ & $\mathbf{\bar{U}}$ & $\mathbf{\Pi_5}$ & $\mathbf{\bar{Q}_7}$ \\
    & $\mathbf{\bar{D}}$ & $\mathbf{H}$ & $\mathbf{\bar{D}}$ & $\mathbf{S}$ \\
    & $\mathbf{\bar{Q}_1}$ & $\mathbf{S}$ & $\mathbf{\bar{Q}_1}$ & $\mathbf{H^{\dagger}}$ \\
    & $\mathbf{\bar{Q}_1}$ & $\mathbf{H^{\dagger}}$ & $\mathbf{D}$ & $\mathbf{H^{\dagger}}$ \\
    & $\mathbf{\bar{Q}_7}$ & $\mathbf{\phi_3^{\dagger}}$ & $\mathbf{X_4}$ & $\mathbf{\phi_3^{\dagger}}$ \\
    & $\Delta_1$ & $\Pi_5$ & $\bar{U}$ & $\Pi_5$ \\
    & $\Delta_3$ & $\omega_5^{\dagger}$ & $\Delta_3$ & $\Pi_{11}$ \\
    & $\omega_5^{\dagger}$ & $\Delta_3$ & $\omega_5^{\dagger}$ & $\bar{N}_2$ \\
    & $\Delta_3$ & $\Pi_{11}$ & $\bar{X}_5$ & $\Pi_{11}$ \\
    & $\Pi_{11}^{\dagger}$ & $N_2$ & $\omega_5$ & $\bar{\Delta}_3$ \\
    & $\Pi_{11}^{\dagger}$ & $N_2$ & $\Pi_{11}^{\dagger}$ & $\bar{\Delta}_3$ \\
    & $\Pi_5^{\dagger}$ & $\bar{E}$ & $\Pi_5^{\dagger}$ & $\bar{\Delta}_1$ \\
    & $Q_7$ & $\Pi_5^{\dagger}$ & $\bar{E}$ & $\Pi_5^{\dagger}$ \\
    & $X_4$ & $\omega_5^{\dagger}$ & $\Delta_3$ & $\omega_5^{\dagger}$ \\
    \hline
    \end{tabular}
    \qquad
    \begin{tabular}{|c|cccc|}
    \hline
    operator & & & & \\
    \hline
    \hline
    $\mathcal{O}_{qu}$ & $\mathbf{\Pi_{13}}$ & $\mathbf{\bar{\Delta}_3}$ & $\mathbf{\Pi_{13}}$ & $\mathbf{N_2}$ \\
    & $\mathbf{\phi_3}$ & $\mathbf{\bar{Q}_5}$ & $\mathbf{\phi_3}$ & $\mathbf{\bar{X}_4}$ \\
    & $\mathbf{\Pi_{11}^{\dagger}}$ & $\mathbf{Q_7}$ & $\mathbf{\Pi_{11}^{\dagger}}$ & $\mathbf{X_5}$ \\
    & $\mathbf{\bar{X}_7}$ & $\mathbf{\Pi_{13}}$ & $\mathbf{\bar{\Delta}_3}$ & $\mathbf{\Pi_{13}}$ \\
    & $\mathbf{N_2}$ & $\mathbf{\Pi_{11}^{\dagger}}$ & $\mathbf{Q_7}$ & $\mathbf{\Pi_{11}^{\dagger}}$ \\
    & $\mathbf{H}$ & $\mathbf{\bar{Q}_1}$ & $\mathbf{H}$ & $\mathbf{\bar{D}}$ \\
    & $\mathbf{U}$ & $\mathbf{S}$ & $\mathbf{U}$ & $\mathbf{H}$ \\
    & $\mathbf{Q_1}$ & $\mathbf{H^{\dagger}}$ & $\mathbf{Q_1}$ & $\mathbf{S}$ \\
    & $\mathbf{X_5}$ & $\mathbf{\phi_3}$ & $\mathbf{\bar{Q}_5}$ & $\mathbf{\phi_3}$ \\
    & $\mathbf{S}$ & $\mathbf{\bar{U}}$ & $\mathbf{S}$ & $\mathbf{\bar{Q}_1}$ \\
    & $\mathbf{H^{\dagger}}$ & $\mathbf{\bar{Q}_7}$ & $\mathbf{H^{\dagger}}$ & $\mathbf{\bar{U}}$ \\
    & $\mathbf{U}$ & $\mathbf{S}$ & $\mathbf{\bar{Q}_1}$ & $\mathbf{H}$ \\
    & $\mathbf{U}$ & $\mathbf{H}$ & $\mathbf{\bar{Q}_1}$ & $\mathbf{H}$ \\
    & $\mathbf{D}$ & $\mathbf{H^{\dagger}}$ & $\mathbf{\bar{Q}_7}$ & $\mathbf{H^{\dagger}}$ \\
    
    & $\mathbf{\bar{U}}$ & $\mathbf{S}$ & $\mathbf{\bar{U}}$ & $\mathbf{\Pi_5}$ \\
    & $\Delta_3$ & $\omega_5^{\dagger}$ & $X_7$ & $\omega_5^{\dagger}$\\
    & $\omega_5$ & $\bar{E}$ & $\omega_5$ & $\bar{\Delta}_3$ \\
    & $\bar{X}_7$ & $\omega_5$ & $\bar{\Delta}_3$ & $\Pi_{13}$ \\
    & $\omega_5$ & $\bar{E}$ & $\Pi_5^{\dagger}$ & $\bar{\Delta}_3$ \\
    & $\Pi_5^{\dagger}$ & $Q_1$ & $\Pi_5^{\dagger}$ & $U$ \\
    & $\bar{Q}_{11}$ & $\omega_5$ & $\bar{E}$ & $\omega_5$ \\
    & $\bar{E}$ & $\Pi_5^{\dagger}$ & $Q_1$ & $\Pi_5^{\dagger}$ \\
    & $\bar{E}$ & $\omega_5$ & $\bar{E}$ & $\Pi_5^{\dagger}$ \\
    & $Q_1$ & $\Pi_5^{\dagger}$ & $Q_1$ & $S$ \\
    & $\Pi_5^{\dagger}$ & $Q_1$ & $S$ & $U$ \\
    & $\Pi_5$ & $\Delta_3$ & $\Pi_5$ & $E$ \\
    & $\Delta_3$ & $\Pi_{13}^{\dagger}$ & $\Delta_3$ & $\omega_5^{\dagger}$ \\
    & $\bar{U}$ & $\Pi_5$ & $\Delta_3$ & $\Pi_5$ \\
    & $\Delta_3$ & $\Pi_5$ & $\Delta_3$ & $\omega_5^{\dagger}$ \\
    \hline
    \end{tabular}
    \caption{Quark specific models, part 1. Boldface models are quark-specific.}
    \label{tab:quark_specific_models_1}
\end{table}

\begin{table}[ht]
\centering
    \begin{tabular}{|c|cccc|}
    \hline
    operator & & & & \\
    \hline
    \hline
    $\mathcal{O}_{uu}$ & $\mathbf{\Pi_{13}}$ & $\mathbf{\bar{\Delta}_3}$ & $\mathbf{\Pi_{13}}$ & $\mathbf{\bar{Q}_{17}}$ \\
    & $\mathbf{\phi_3}$ & $\mathbf{\bar{Q}_5}$ & $\mathbf{\phi_3}$ & $\mathbf{Q_{13}}$ \\
    & $\mathbf{\Pi_{11}^{\dagger}}$ & $\mathbf{Q_7}$ & $\mathbf{\Pi_{11}^{\dagger}}$ & $\mathbf{\Delta_5}$ \\
    & $\mathbf{Q_7}$ & $\mathbf{H}$ & $\mathbf{Q_7}$ & $\mathbf{\Pi_{11}^{\dagger}}$ \\
    & $\mathbf{\bar{\Delta}_3}$ & $\mathbf{\Pi_{13}}$ & $\mathbf{\bar{\Delta}_3}$ & $\mathbf{\Pi_{13}}$ \\
    & $\mathbf{Q_7}$ & $\mathbf{\Pi_{11}^{\dagger}}$ & $\mathbf{Q_7}$ & $\mathbf{\Pi_{11}^{\dagger}}$ \\
    & $\mathbf{H}$ & $\mathbf{\bar{Q}_1}$ & $\mathbf{H}$ & $\mathbf{Q_7}$ \\
    & $\mathbf{\bar{Q}_5}$ & $\mathbf{\phi_3}$ & $\mathbf{\bar{Q}_5}$ & $\mathbf{\phi_3}$\\
    & $\mathbf{S}$ & $\mathbf{\bar{U}}$ & $\mathbf{S}$ & $\mathbf{U}$ \\
    & $\mathbf{\bar{Q}_1}$ & $\mathbf{H}$ & $\mathbf{\bar{Q}_1}$ & $\mathbf{H}$ \\
    & $\mathbf{\bar{Q}_7}$ & $\mathbf{H^{\dagger}}$ & $\mathbf{\bar{Q}_7}$ & $\mathbf{H^{\dagger}}$ \\
    & $\omega_5$ & $\bar{E}$ & $\omega_5$ & $\bar{X}_7$ \\
    & $\bar{\Delta}_3$ & $\Pi_5^{\dagger}$ & $\bar{\Delta}_3$ & $\Pi_{13}$ \\
    & $\Pi_5^{\dagger}$ & $Q_1$ & $\Pi_5^{\dagger}$ & $\bar{\Delta}_3$ \\
    & $Q_1$ & $H^{\dagger}$ & $Q_1$ & $\Pi_5^{\dagger}$ \\
    & $\bar{E}$ & $\omega_5$ & $\bar{E}$ & $\omega_5$ \\
    & $Q_1$ & $\Pi_5^{\dagger}$ & $Q_1$ & $\Pi_5^{\dagger}$ \\
    & $\Delta_3$ & $\Pi_5$ & $\Delta_3$ & $\Pi_5$ \\
    \hline
        $\mathcal{O}_{dd}$ & $\mathbf{\phi_3}$ & $\mathbf{\bar{Q}_{11}}$ & $\mathbf{\phi_3}$ & $\mathbf{Q_7}$ \\
    & $\mathbf{Q_7}$ & $\mathbf{\phi_3}$ & $\mathbf{Q_7}$ & $\mathbf{\Pi_5^{\dagger}}$ \\
    & $\mathbf{Q_7}$ & $\mathbf{\Pi_5^{\dagger}}$ & $\mathbf{Q_7}$ & $\mathbf{\Pi_5^{\dagger}}$ \\
    & $\mathbf{H}$ & $\mathbf{\bar{Q}_5}$ & $\mathbf{H}$ & $\mathbf{Q_1}$ \\
    & $\mathbf{\bar{Q}_5}$ & $\mathbf{H}$ & $\mathbf{\bar{Q}_5}$ & $\mathbf{H}$ \\
    & $\mathbf{S}$ & $\mathbf{\bar{D}}$ & $\mathbf{S}$ & $\mathbf{D}$ \\
    & $\mathbf{\bar{Q}_1}$ & $\mathbf{H^{\dagger}}$ & $\mathbf{\bar{Q}_1}$ & $\mathbf{H^{\dagger}}$ \\
    & $\mathbf{\bar{Q}_7}$ & $\mathbf{\phi_3^{\dagger}}$ & $\mathbf{\bar{Q}_7}$ & $\mathbf{\phi_3^{\dagger}}$ \\
    & $\Pi_{11}^{\dagger}$ & $Q_{13}$ & $\Pi_{11}^{\dagger}$ & $\bar{\Delta}_3$ \\
    & $\Pi_5^{\dagger}$ & $Q_7$ & $\Pi_5^{\dagger}$ & $\bar{\Delta}_1$ \\
    & $\Delta_1$ & $\Pi_5$ & $\Delta_1$ & $\Pi_5$ \\
    & $\Delta_3$ & $\Pi_{11}$ & $\Delta_3$ & $\Pi_{11}$ \\
    \hline
    \end{tabular}
    \qquad
    \begin{tabular}{|c|cccc|}
    \hline
    operator & & & & \\
    \hline
    \hline
    $O_{ud}$ & $\mathbf{\Pi_{13}}$ & $\mathbf{\bar{\Delta}_3}$ & $\mathbf{\Pi_{13}}$ & $\mathbf{\bar{Q}_{11}}$ \\
    & $\mathbf{\phi_3}$ & $\mathbf{\bar{Q}_5}$ & $\mathbf{\phi_3}$ & $\mathbf{Q_7}$ \\
    & $\mathbf{Q_{13}}$ & $\mathbf{\phi_3}$ & $\mathbf{Q_7}$ & $\mathbf{\Pi_{11}^{\dagger}}$ \\
    & $\mathbf{Q_7}$ & $\mathbf{H}$ & $\mathbf{Q_7}$ & $\mathbf{\Pi_{5}^{\dagger}}$ \\
    & $\mathbf{\Delta_5}$ & $\mathbf{\Pi_{13}}$ & $\mathbf{\bar{\Delta}_3}$ & $\mathbf{\Pi_{13}}$ \\
    & $\mathbf{Q_{13}}$ & $\mathbf{\Pi_{11}^{\dagger}}$ & $\mathbf{Q_7}$ & $\mathbf{\Pi_{11}^{\dagger}}$\\
    & $\mathbf{Q_7}$ & $\mathbf{\Pi_{11}^{\dagger}}$ & $\mathbf{Q_7}$ & $\mathbf{\Pi_5^{\dagger}}$\\
    & $\mathbf{\phi_3}$ & $\mathbf{\bar{Q}_5}$ & $\mathbf{H}$ & $\mathbf{Q_7}$ \\
    & $\mathbf{H}$ & $\mathbf{\bar{Q}_1}$ & $\mathbf{H}$ & $\mathbf{Q_1}$ \\
    & $\mathbf{\bar{Q}_{11}}$ & $\mathbf{\phi_3}$ & $\mathbf{\bar{Q}_5}$ & $\mathbf{\phi_3}$ \\
    & $\mathbf{\bar{Q}_5}$ & $\mathbf{\phi_3}$ & $\mathbf{\bar{Q}_5}$ & $\mathbf{H}$ \\
    & $\mathbf{S}$ & $\mathbf{\bar{U}}$ & $\mathbf{S}$ & $\mathbf{D}$ \\
    & $\mathbf{H^{\dagger}}$ & $\mathbf{\bar{Q}_7}$ & $\mathbf{H^{\dagger}}$ & $\mathbf{Q_5}$ \\
    & $\mathbf{\bar{Q}_5}$ & $\mathbf{H}$ & $\mathbf{\bar{Q}_1}$ & $\mathbf{H}$ \\
    & $\mathbf{\bar{Q}_7}$ & $\mathbf{\Pi_{11}}$ & $\mathbf{\bar{Q}_7}$ & $\mathbf{\phi_3^{\dagger}}$ \\
    & $\mathbf{\bar{Q}_1}$ & $\mathbf{H^{\dagger}}$ & $\mathbf{\bar{Q}_7}$ & $\mathbf{H^{\dagger}}$ \\
    & $\mathbf{\bar{Q}_7}$ & $\mathbf{H^{\dagger}}$ & $\mathbf{\bar{Q}_7}$ & $\mathbf{\phi_3^{\dagger}}$ \\
    & $\mathbf{\bar{Q}_7}$ & $\mathbf{\phi_3^{\dagger}}$ & $\mathbf{\bar{Q}_{13}}$ & $\mathbf{\phi_3^{\dagger}}$ \\
    & $\omega_5$ & $\bar{E}$ & $\omega_5$ & $\bar{X}_4$ \\
    & $\Pi_{11}^{\dagger}$ & $Q_7$ & $\Pi_{11}^{\dagger}$ & $\bar{\Delta}_3$ \\
    & $\Delta_5$ & $\Pi_{11}^{\dagger}$ & $\bar{\Delta}_3$ & $\Pi_{13}$ \\
    & $\Pi_{11}^{\dagger}$ & $Q_7$ & $\Pi_5^{\dagger}$ & $\bar{\Delta}_3$ \\
    & $\Pi_5^{\dagger}$ & $Q_1$ & $\Pi_5^{\dagger}$ & $\Delta_1^{\dagger}$ \\
     & $Q_7$ & $H$ & $Q_1$ & $\Pi_5^{\dagger}$ \\
    & $N_2$ & $\omega_5$ & $\bar{E}$ & $\omega_5$ \\
    & $Q_7$ & $\Pi_5^{\dagger}$ & $Q_1$ & $\Pi_5^{\dagger}$ \\
    & $\Pi_5$ & $\Delta_3$ & $\Pi_5$ & $\bar{Q}_7$ \\
    & $\bar{Q}_1$ & $\Pi_5$ & $\bar{Q}_1$ & $H^{\dagger}$ \\
    & $\Delta_3$ & $\Pi_{13}^{\dagger}$ & $\Delta_3$ & $\Pi_{11}$ \\
    & $\Delta_1$ & $\Pi_5$ & $\Delta_3$ & $\Pi_5$ \\
    & $\Delta_3$ & $\Pi_5$ & $\Delta_3$ & $\Pi_{11}$ \\
    & $\Delta_3$ & $\Pi_{11}$ & $\bar{\Delta}_5$ & $\Pi_{11}$ \\
    \hline
    \end{tabular}
    \caption{Quark specific models, part 2. Boldface models are quark-specific.}
    \label{tab:quark_specific_models_2}
\end{table}

\bibliographystyle{JHEP}
\bibliography{paper_exits}

\end{document}